\documentclass{emulateapj}
\usepackage{color}
\usepackage[below]{placeins}
\usepackage{natbib}
\slugcomment{accepted by the Astrophysical Journal}

\def\gtorder{\mathrel{\raise.3ex\hbox{$>$}\mkern-14mu
    \lower0.6ex\hbox{$\sim$}}}
\def\ltorder{\mathrel{\raise.3ex\hbox{$<$}\mkern-14mu
    \lower0.6ex\hbox{$\sim$}}}
\def\hmpc{h^{-1} \mathrm{Mpc}}

\def\msun{M_\odot} 
\def\hmsun{h^{-1} M_\odot}
\def \etal {{\it et al.}}

\def \ie {{\it i.e.,} }
\def \eg{{\it e.g.,}}

\def \br{{\bf r}}
\def \bk{{\bf k}}
\def\rd{{\rm d}}

\newcommand {\nbody} {$N$-body}
\newcommand {\at} {$a$}
\newcommand {\cs} {$c$}
\newcommand {\rs} {$R_{\rm s}$}
\newcommand {\ms} {$M_{\rm s}$}
\newcommand {\ks} {$K_{\rm s}$}
\newcommand {\rvir} {$R_{\rm vir}$}
\newcommand {\mvir} {$M_{\rm vir}$}
\newcommand {\kvir} {$K_{\rm vir}$}
\newcommand {\rhos} {$\rho_{\rm s}$}
\newcommand {\phis} {$\phi_{\rm s}$}
\newcommand {\phiv} {$\phi_{\rm vir}$}
\newcommand {\qs} {$Q_{\rm s}$}
\newcommand {\qvir} {$Q_{\rm vir}$}
\newcommand {\ls} {$\lambda_{\rm s}$}
\newcommand {\lvir} {$\lambda_{\rm vir}$}

\shorttitle{Evolution of dark matter halo density profiles}
\shortauthors{Romano-D\'{\i}az et al.}

\begin{document}

\title{Evolution of characteristic quantities for dark 
matter halo density profiles}

\author{
Emilio Romano-D\'{\i}az\altaffilmark{1,2},
Yehuda Hoffman\altaffilmark{1},
Clayton Heller\altaffilmark{3},
Andreas Faltenbacher\altaffilmark{4},
Daniel Jones\altaffilmark{2}\\ 
and Isaac Shlosman\altaffilmark{2}
}
\altaffiltext{1}{
Racah Institute of Physics, Hebrew University; Jerusalem 91904, Israel
}
\altaffiltext{2}{
Department of Physics and Astronomy, 
University of Kentucky, 
Lexington, KY 40506-0055, 
USA
}
\altaffiltext{3}{
Department of Physics, 
Georgia Southern University, 
Statesboro, GA 30460, 
USA
}
\altaffiltext{4}{
Physics Department, 
University of California, 
Santa Cruz CA 95064, 
USA
}

\begin{abstract}

We have investigated the effect of an assembly history on the
evolution of galactic dark matter (DM) halos of $10^{12} \hmsun$ using
Constrained Realizations of random Gaussian fields. Five different
realizations of a DM halo with distinct merging histories were
constructed and have been evolved using collisionless high-resolution
\nbody\ simulations. Our main results are: A halo evolves {\it via} a
sequence of quiescent phases of a slow mass accretion intermitted by
violent episodes of major mergers. In the quiescent phases, the
density is well fitted by an NFW profile, the inner scale radius \rs\
and the mass enclosed within it remain constant, and the virial radius
(\rvir ) grows linearly with the expansion parameter $a$. Within
each quiescent phase the concentration parameter ($c$) scales as $a$,
and the mass accretion history (\mvir) is well described by the
Tasitsiomi \etal\ fitting formula. In the violent phases the halos are
not in a virial dynamical equilibrium and both \rs\ and \rvir\ grow
discontinuously.  The violent episodes drive the halos from one NFW
dynamical equilibrium to another. The final structure of a halo,
including $c$, depends on the degree of violence of the major mergers
and the number of violent events. Next, we find a distinct difference
between the behavior of various NFW parameters taken as averages over
an ensemble of halos and those of individual halos. Moreover, the
simple scaling relations $c-$\mvir\ do not apply to the entire
evolution of individual halos, and so is the common notion that late
forming halos are less concentrated than early forming ones. The
entire evolution of the halo cannot be fitted by single analytical
expressions.

\end{abstract}



\section{Introduction}
\label{sec:intro}

The hierarchical buildup of cold dark matter (CDM) halos is a strongly
nonlinear process. It is associated with the buildup of a unique
density profile --- one of the most fundamental characteristics of the
DM halos. Based on \nbody\ simulations, \citet[hereafter NFW]{nfw97}
found that the CDM halos can be universally fitted by a two-parameter
functional form:
\begin{equation}
\rho(r) = \frac{4 \rho_{\rm s}}{(r/R_{\rm s})(1+r/R_{\rm s})^2} \,,
\label{eq:nfw}
\end{equation}
where \rs\ is a characteristic ``inner" radius at which the
logarithmic density slope is $-2$ and \rhos\ is the density at \rs.
The cosmological evolution of the NFW parameters, therefore, is an
issue of a broad interest and is addressed in this work.

A useful alternative parameter to describe the shape of the profile is
the so-called halo concentration parameter defined as $c = R_{\rm
vir}/R_{\rm s}$, where \rvir\ is the halo virial radius.  Although
such a profile has been confirmed by numerous \nbody\ simulations, the
exact value of the inner slope parameter is uncertain. Several authors
have found that halos have density cusps steeper
\citep[\eg][]{fm97,fm03,moore99,ghigna00} or shallower
\citep[\eg][]{sco00,tn01} than $-1$, the NFW value. The NFW profile
and its modifications are specific cases of a three-parameter profile
family proposed by \cite{h90}, and further developed by \cite{zhao96}.
On the other hand, \citep{js00,js02}, \cite{kkbp01} and
\cite{ricotti03} found that the CDM halos do not maintain the
universal density profiles, and the inner slope changes from
galaxy-size halos to cluster-size halos \citep[see also][]{t04}.
 
These numerical results appear in conflict with the observational
evidence --- rotation curves of low surface brightness galaxies yield
density profiles with nearly constant density {\it cores}
\citep[\eg][]{fp94,sb00}. Studies of brighter galaxies imply similar
problems \citep{sb00,de-Blok:2002aa, gen04}.  While some studies of
the weak lensing seem to support the NFW density profile
\citep[\eg][]{Hoekstra:2004aa}, in a specific case of the cluster
A1689, the required concentration parameter is dramatically larger
than the typical value obtained in the simulation of a cluster-size
object \citep{broadhurst05}.

In principle, one can list three different options to resolve the
above difficulties. First, the inner few-kpc rotation curves of disk
galaxies can be contaminated with the (unresolved) {\it non}-circular
motions triggered \eg\ by stellar bars
\citep[\eg][]{blais01,bolatto02,simon03,weldrake03,simon05}.  Second,
the NFW approach is to neglect the triaxial nature of DM halos by
``sphericalizing'' while analyzing them.  Furthermore, the issue of
the ``cusps'' might be related to the effective resolution in \nbody\
simulations or other numerical effects \citep[\eg][]{moore98}.

Finally, the difference between the collisionless simulations and
observations can underscore a physical effect, like the absence of
dissipation in the pure DM models and the presence of dissipative
baryons in the real galaxies. This apparent discrepancy can be
reconciled within the context of the CDM model by considering the
effect of clumpy baryons erasing the cusps on relevant spatial scales,
from galactic halos to clusters of galaxies
\citep{El-Zant:2001aa,El-Zant:2004aa}.  Alternatively, it has been
suggested that the stellar bars facilitate the destruction of DM cusps
\citep{Weinberg:2002aa}, but this has been disputed by \citet{dehn05}.

There has been no ``natural" explanation for the origin and
``universality" of the NFW profile. While analytical models
necessarily invoke spherical symmetry, numerical simulations emphasize
the background asymmetry. \citet{gg72} proposed an analytical model
which invokes the collapse of uniform spherical perturbations of
collisionless DM in an expanding universe. This model explains some of
the global properties of virialized halos (\eg\ mean density and
size), but does not account for the density profile.  A more rigorous
and exact analytical solution relevant to the problem is that of a
single scale free spherical density perturbation in a Friedmann
universe, the so-called secondary infall model \citep{gunn77, fg84,
bert85} and its application to cosmological models
\citep{hs85,rg87,zh93,lh00}.  An orthogonal approach has been
suggested to explain the emergence of the NFW density profile as the
outcome of mergers between substructures and progenitors of halos
\citep{sw98,ns99,sco00}. \citet{ez05} has shown that the NFW density
profiles remain invariant under interactions with the subhalos --- a
necessary step toward the universality of this mass distribution.

Several authors \citep[\eg][]{nusser01,ascasibar04,Hiotelis:2002aa}
extended the secondary infall model by allowing for non-radial
orbits. They have shown that by properly tuning the anisotropy of the
orbits the model yields a cuspy density profile, quite similar to the
NFW one -- a task performed by major mergers \citep{Lu:2006aa}.
  
At least in spherically-symmetric models, it was argued that the halo
concentration $c$ increases with lower mass because they form early,
when the universe density is higher. This leads to higher \rhos\
\citep[\eg][]{ens01}. On the other hand, within the hierarchical
structure formation, halos of a particular mass might form at
different times, so their mass is not a unique function of their
formation time, but can depend on their environment.  Hence, the
central densities do not necessarily reflect the properties of the
universe at specific times \citep{avila05,w05}, but the significance
of this effect is not clear yet.

\citet[hereafter, W02]{w02} and others have provided simple analytical
fits to the evolution of various quantities which characterize the
halos, \eg\ $c$, halo formation time, its mass, \rs, etc. Yet, the
scatter around these fits is considerable and its origin is
unclear. This has led us to embark on a series of numerical
experiments carefully designed to investigate the evolution of the
basic halo parameters.  By using constrained realizations (CRs;
\citep{hr91}) of the initial density field we can `design' the merging
history of a single galactic halo.

Most of the studies of the cosmological evolution of DM halos have
focused analyzing large ensembles of simulated halos and studying
their ensemble averaged properties \citep[\eg\
W02,][]{lk99,bullock01ck,bullock01sf,peirani04,avila05,Reed:2005aa,Shaw:2005ua}.
Closer inspection of individual halos shows that their evolution is
not smooth and monotonic but goes through a number of
discontinuities. The question that arises is whether this erratic
behavior has a numerical origin (e.g., numerical resolution and coarse
time sampling) or is a consequence of the underlying physics. In the
latter case, this implies that the smooth fitting formulae do not
provide a good approximation to the dynamics and evolution of
individual halos. This issue is addressed here.

In a previous work (\cite{erd06}, hereafter Paper I), we have used the
CRs of a single galactic halo to show that its evolution can be
described by a series of step functions.  By changing the merging
history in the initial conditions of the same single galactic halo, we
have found that the inner structure (\ie\ within \rs) remains
unchanged in the slow accretion phase and it evolves violently in the
fast accretion phase.

In the present paper, we extend and elaborate our previous results
from Paper~I and analyze additional physical quantities such as the shape,
angular momentum, mass, kinetic and potential energies of halos. Our
results are developed in the context of the Open Cold Dark Matter
(OCDM) scenario. However, our main conclusions are independent of the
exact cosmological model under consideration.  Since we are interested
in the relative differences between the various simulated halos, we
shall assume here that the NFW density profile is a good enough
approximation to a simulated halo density profile.

The present paper is structured as following.  The numerical
experiments are described in \S~\ref{sec:numerics}. The mass accretion
history of the primary halo is presented in \S ~ \ref{sec:mahs} and
our analysis of the resulting halos is described in \S ~
\ref{sec:nfw}.  Additional analysis of the structure of the DM halos
is presented in \S~\ref{sec:results} and a general discussion in
\S~\ref{sec:disc}.


\section{Numerical experiments}
\label{sec:numerics}

\begin{figure*}[!t]
\epsscale{1.}
\plotone{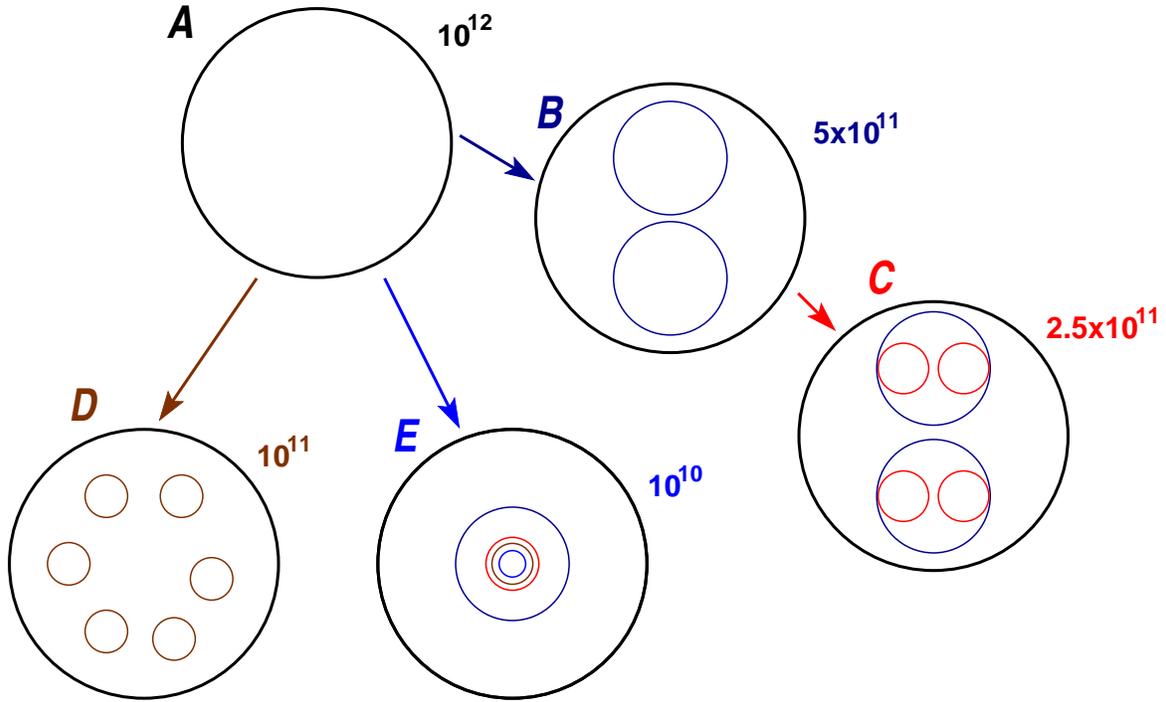}
\caption{Model configurations. Schematic representation of our five 
models and their corresponding constraints. The top right labels
indicate the mass of the constraint. The color of the labels indicate
the constraint level within each circle.}
\label{fig:models}
\end{figure*}

The common approach to study the evolution of single halos with high
numerical resolution and from a cosmological point of view consists of
two steps.  First, halos of interest are identified in a large
cosmological \nbody\ simulation. Second, the particles which make up
these halos are traced back to the initial conditions. The region
enclosing these particles is then re-simulated with a higher
resolution.  An alternative approach is to find a way of producing the
requested structures in agreement with the cosmological model
imposed. This can be done by setting up the initial density fields by
CRs of Gaussian random fields, following the prescription of
\cite{hr91}.

The use of CRs to set up the initial conditions as an input for the
cosmological simulations has been applied to study several aspects of a
structure formation.  The CRs are also instrumental in using the data to
set up the cosmological simulations. This has been done for studies
related to the matter distribution in the nearby universe
\citep[\eg,][]{bh98,kkh02,erd04}. However, none of these studies have
focused on the evolution of single galactic halos. Below we describe
the main characteristics of our CRs, while a mathematical description
of the CR formalism is exposed in Appendix A.

\subsubsection{The models}
\label{sec:models}

We have designed a set of five different models, \ie\ experiments, to
probe different merging histories of a few times $10^{12}\hmsun$ halo
in an OCDM cosmology. The halo is constrained to have various
substructure on different mass scales and locations, designed to
collapse at different times. The spherical top-hat model is used here
to set the numerical value of the constraints. The model provides the
collapse time of the substructures as a function of the initial
density. This is used only as a general and rough guide because the
substructures are neither spherical nor isolated and hence their
collapsing time might vary. Furthermore, the constraints used here do
not control the experiments fully. The nonlinear dynamics can, in
principle, affect the evolution in a way not fully anticipated from
the initial conditions. Even more important is the role of the random
component in the CRs. Thus depending on the nature of the constraints
and the power spectrum assumed, the random component can form other
significant substructures at different locations and mass scales. This
can be handled by adding more constraints and varying their numerical
values. The price to pay is that the modeled entity will be more
``synthetic.''

\begin{deluxetable}{ccccccc}
\tablewidth{0pt}
\tablehead{
\colhead{Model}  &  \multicolumn{5}{c}{Constraints ($\hmsun$)}  &  
\colhead{$z_{\rm coll}$}\\
& $10^{12}$ & $5\times 10^{11}$ & $2.5\times 10^{11}$ & $10^{11}$
& $10^{10}$ & }
\startdata
	A & 1 & & & & & 2.1\\
	B & 1 & 2 & &  & & 3.7\\
	C & 1 & 2 & 4 & &  & 5.7\\
	D & 1 & & & 6 & & 7.0\\
	E & 1 & 1& 1 & 1 & 1 & 8.9\\
\enddata
\tablecomments{\tabletypesize{\small}Characteristics of the CR models:
the first column indicates the model labels. The second, third ---
sixth columns indicate the level of the mass constraints expressed in
$\hmpc$. The given numbers represent the number of constraints imposed at
each level for each model. The last column indicates the collapse
redshift ($z_{\rm coll}$) of the smallest (in mass) constraint imposed
on each model. \label{table:models}}
\end{deluxetable}

Table~\ref{table:models} presents the main characteristics of our five
models.  The first column indicates the label of the model. The
second-to-sixth columns show the masses of the different constraints
(expressed in $\hmsun$). The last column states the collapse redshift
of the last/smallest constraint in each configuration.  All five
models are embedded in a region corresponding to a mass of $\sim
10^{13}\hmsun$ in which the over-density is constrained to be zero ---
a region of an unperturbed Friedmann universe. This spatial scale is
larger by about a factor of three (in mass) than the size of the
computational sphere. Therefore, this constraint cannot be exactly
fulfilled, yet it constrains the large scale modes of the realizations
to obey it. Model A --- our benchmark model, is based only on one
constraint of $10^{12}\hmsun$. All other models have substructures
imposed onto this constraint as illustrated in
Figure~\ref{fig:models}.  Each imposed constraint level is indicated
by a different color and by its mass at the top right part
of model. The position of the constraints is schematic and only serves
to get a visual impression of the models. The small-scale constraints
imposed over our benchmark model are aimed to modify the merging
history of the benchmark model. They change the collapse time as well
as the number of major mergers that each model will pass through. In
the hypothetical case, when the random part of the CR method does not
contribute to the major merger history of the models, model B will
experience only one major merger, while model C will have two, and
model D can have more than three major mergers. Model E is aimed to
mimic a radial collapse by imposing concentric (nested) constraints.

All models have been constructed with the same random seed.  This
guarantees that the linear external matter distribution and the random
component will be the same for all models \citep{hr91,vdw96}.  All the
density constraints constitute $(2.5 - 3.5)\sigma$ perturbations,
where $\sigma^2$ is the variance of the appropriately smoothed field.
The constraints were imposed on a box size of $4~\hmpc$. In order to
avoid the edge effects with the smoothing procedure, the constrained
box has been immersed in a $8~\hmpc$ box on a cubic grid with 256 grid
cells per dimension. The constrained inner box was carved out and
resulted in a box size of $4~\hmpc$ and 128 grid cells per dimension.

\subsection{N-body simulations}
\label{sec:nbody}

Once the constrained initial conditions have been generated, we have
applied the Zel'dovich formalism \citep{zeldovich} in order to obtain
the initial positions and velocity displacements at redshift $z=120$,
using physical rather than comoving coordinates. A sphere of $2~\hmpc$
(comoving) radius was carved out from the Zel'dovich evolved
fields. All our constraints are totally embedded within this maximum
radius sphere. The total number of particles selected within this
volume is $\sim 1.3 \times 10^6$, resulting in a mass resolution of
$3.6\times10^{6} \msun$. The gravitational softening is 500~pc.

We have evolved each linear density field into the non-linear regime,
until $z=0$ (present time), using the updated FTM-4.4 version of our
hybrid code \citep{hel94,hel95} with the {\tt falcON} routine
\citep{dehn02}. The code is about ten times faster than the optimally
coded \citet{Barnes:1986aa} tree code.  The equations of motion in the
FTM-4.4 include the term describing the expansion of the universe.
The code has been successfully tested against the Santa Barbara
Cluster model \citep{sta-barbara} (see Appendix~B).

Because we are interested in following the merging history of our
models as close as possible, we have sampled the dynamical evolution
of the systems with 165 time outputs spaced linearly with the
expansion parameter $a$. Each halo is resolved at $z=0$ with around
$1.1\times 10^6$ particles within the virial radius.

\section{Mass Accretion Histories}
\label{sec:mahs}

Halos grow according to a hierarchical formation scenario, in which
they are assembled via merging of different mass and size halos,
together with a more gentle accretion. Several studies have found that
the mass accretion history of halos (MAH) may affect the various halo
parameters, including \rs, while maintaining their NFW density
profiles \citep[W02]{nfw97,bullock01sf,z03b,t04}. Therefore, it is
important to analyze the MAHs of our five models.

W02 proposed to fit the MAHs of halos by an exponential function in
the form:
\begin{equation}
M(a) = M_0 ~e^{-\alpha(1/a-1)}
\label{eq:mahw}
\end{equation}
where $a$ is the expansion factor and $M_0$ is the final virial halo
mass. \cite{vdb02} arrived to a very similar expression with a
two-parameter function using the extended Press-Schechter
formalism. \cite{t04} found that in many cases Eq.~\ref{eq:mahw}
represents a poor fit to the individual MAHs halos, in particular when
halos experience an intense and violent activity, up to the present
time. \citeauthor{t04} proposed a more general MAH fit of the form:
\begin{equation}
M(a) = a^p ~M_0 ~e^{-\alpha(1/a-1)}
\label{eq:maht}
\end{equation}
where $p=0$ reduces to the W02 model. Despite the improvement, such a smooth fitting
function cannot reproduce all the features in the halo evolution,
especially the major mergers.


\subsection{Halo identification}
\label{sec:halos}

Groups and subgroups identification is not a trivial task. Several
methods have been proposed which use different ``bounding'' criteria,
e.g., Friends of Friends \citep{defw85}, DENMAX \citep{denmax}, HOP
\citep{hop}, SUBFIND \citep{subfind}, etc. Different methods are used
in order to isolate the halos and subhalos within the halos.  We are
interested in following the evolution of the main halo through the
major branch along its merging tree (see \S ~ \ref{sec:trees}).  For
these purposes we use the publically available code
HOP\footnote{available at
http://cmb.as.arizona.edu/~eisenste/hop/hop.html} \citep{hop} which
computes groups based on the isodensity criteria. Next, we identify
the halo centers with the densest particle within the particular
halo. Once the centers are found, we grow the halos by adding thin
spherical shells and computing the total density until it reaches the
value of $\Delta_{\rm c}$ times the mean density of the universe.
Therefore, a halo within the virial radius will be specified as a
function of redshift as:
\begin{equation}
M_{\rm vir}(z) = \frac{4 \pi}{3} \Delta_{\rm c}(z)\ \rho_{\rm b}(z) 
   \ R_{\rm vir}(z) \,,
\label{eq:mvir}
\end{equation}
where $\Delta_{\rm c}$ is the critical density and it is computed
using the top-hat model, $\rho_{\rm b}$ is the universe density and
$R_{\rm vir}$ is the virial radius of the halo.


\section{Results}
\label{sec:results}

\subsection{Model evolution}

A halo of $\sim 10^{12} \msun$ has formed by $z=0$ in all of our
models, except in model B. At the present time, this model harbors two
halos of $\sim 5\times 10^{11}\msun$ which are on their way to
merge. The final collapse time (i.e., the time of the last major
merger) of the halos is as follows: B, E, A, D and C, from the
earliest to the latest. Defining the halo formation as the time when
half of the mass is in place, results in the same order as above. In
comparison, the top-hat model predicts a different order of collapse
times. This is expected because the random component of the CR method
can introduce other structures of the same level as the
constraints. As a result, the merging history and the collapse time
imposed by these constraints will be modified
\cite[Section~\ref{sec:models}; see also][]{vdw96}. Since we have
warranted that our five models have the same random component, all of
them will be affected in the same way.

\begin{figure*}[!t]
\epsscale{1.0}
\plotone{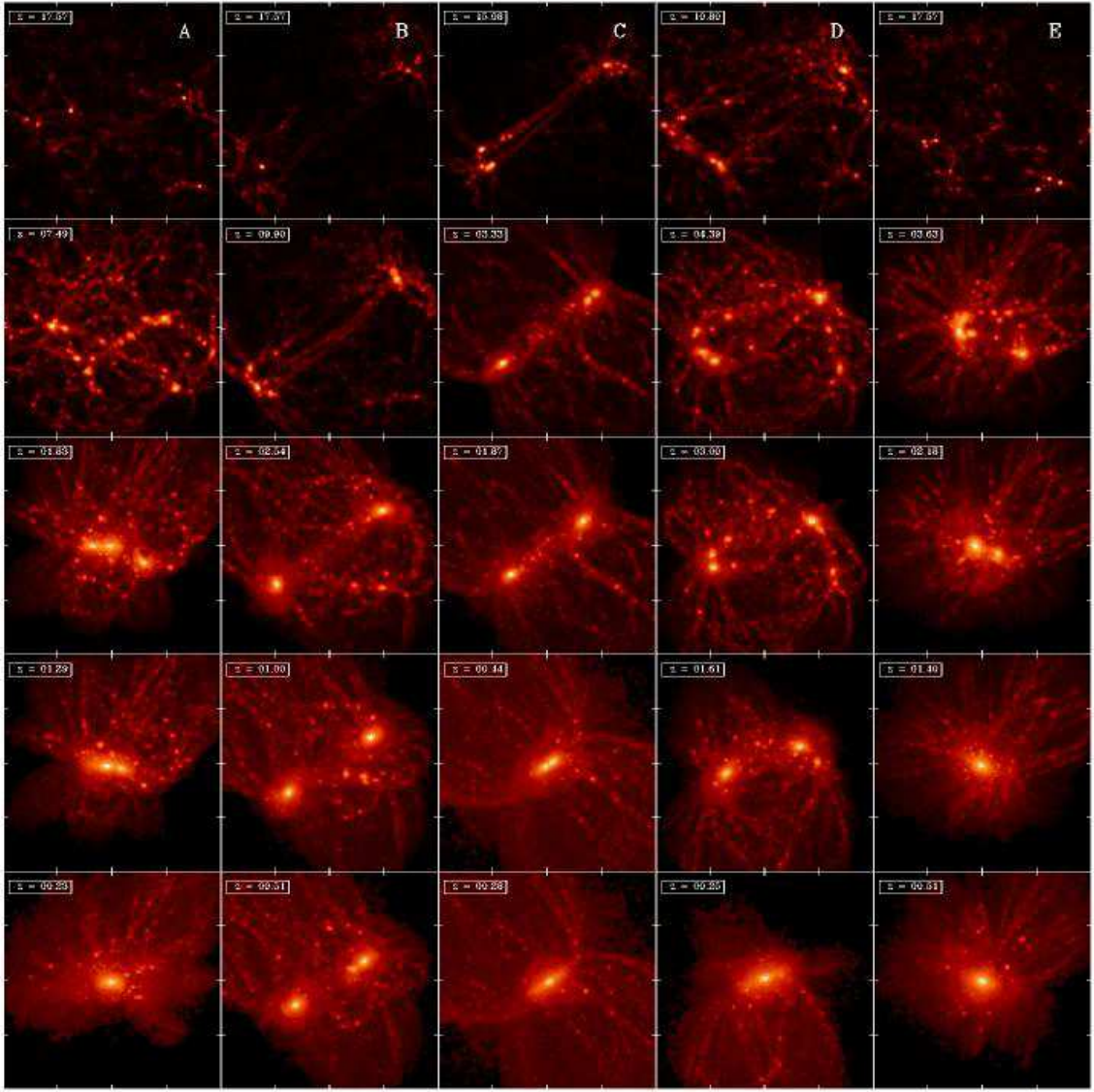}
\caption{Time evolution snapshots for the five models. Models are
arranged from left to right columns: A, B, C, D and E. Each frame
represents the logarithmic density field weighted by the mean density
at the given time. The top left labels in each frame indicate the
redshift of a given snapshot.}
\label{fig:frames}
\end{figure*}

The evolution of the models can be observed in their density
projections (Fig.~\ref{fig:frames}) and the respective merging trees
(Fig.~\ref{fig:trees}). The effect of different constraints imposed on
the halos can be noticed from their evolution on these
diagrams. Figure~\ref{fig:trees} is a schematic representation of the
halo growth and the merging history of the models. The size of the
circles is proportional to the mass (left column numbers) of the
depicted halos (see Section~\ref{sec:trees}). This representation
underscores the effect of the CRs on the evolution of these
models. The model A evolves through three major mergers. Since model B
harbors two very similar halos, we only present the evolution of one
of them (the most massive). It has two nearly simultaneous major
mergers of an exceptional strength, in its early phase. Because the
halo is out of the dynamical equilibrium during this time, we consider
this double merger event as a single one.  Model C frame shows the
evolution of the four constraints, their collapse into two major
entities and formation of one single halo. In the model D, the six
constraints and their development into two major halos and the final
collapse can be clearly followed from the respective tree. The early
formation of the model E through two recognizable major mergers is
neatly shown in its merger tree panel.

Figure~\ref{fig:frames} exhibits density maps in the co-moving
coordinates of the models for five different epochs taken from the
respective merger trees. The first column refers to model A, the
second --- to B, the third --- to C, the fourth --- to D, and the last
one --- to model E.  The color scale is proportional to the logarithm
of the density field. The first row shows the different initial
conditions between the five models. The last row shows the final state
of the halo. Shown are the random frames after the last major merger
that each model experienced, and not the last snapshot of our
simulations. The intermediate frames illustrate different dynamical
epochs for the halos. In model A, the frames depict the moments prior
to two major mergers. Model B frames clearly show the two imposed
constraints --- the system, although gravitationally bound, does not
merge at the present time, but it will do this in the future. Models C
and D show the early collapse of the small constraints, the formation
of the two major clumps and their way to collapse into a single
halo. Model E (the first model to collapse into a $\sim 10^{12} \msun$
halo) shows a more radial filamentary structure around it, a direct
consequence of the imposed constraints. The amount of substructure
changes from model to model, with the model D exhibiting the richest
substructure.

\subsection{Merger trees}
\label{sec:trees}

\begin{figure*}[!t]
\epsscale{1.0}
\plotone{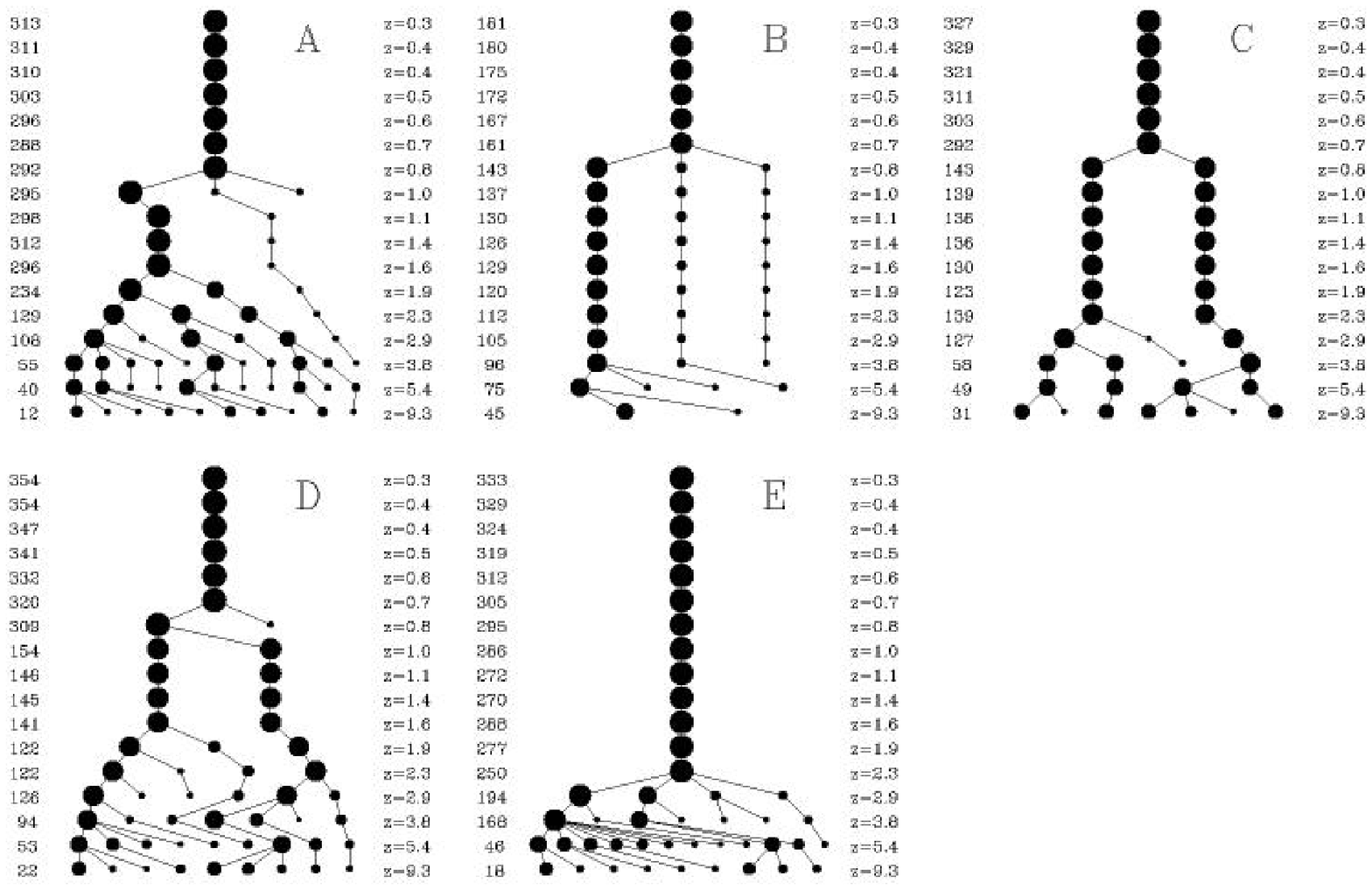}
\caption{Schematic merger trees for our five models. The right columns
indicate the redshift frame. Circles are proportional to the mass of
the halos at the given times.}
\label{fig:trees}
\end{figure*}

\begin{figure*}[!t]
\epsscale{1.0}
\plotone{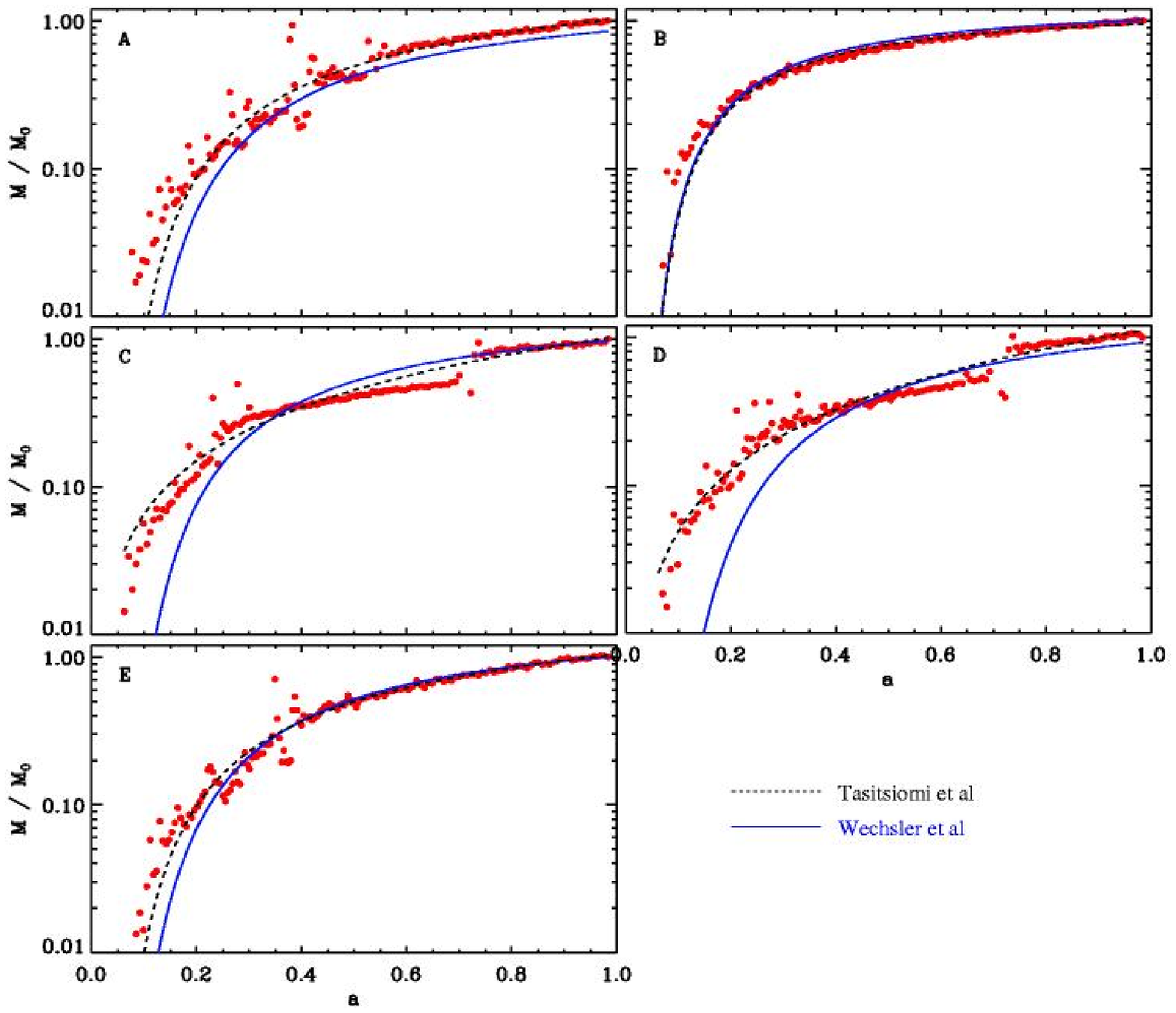}
\caption{Mass Accretion Histories for our five models. The red dots
indicate the measured mass normalized by the final mass of the
respective halo. The blue continuous line represents the best fit 
from the W02 model, while the dashed black lines the \cite{t04} model.}
\label{fig:mah}
\end{figure*}

Since we have identified all halos for basically each of the output
times of our simulations (we have found halos for all models from $z
\approx 20$ onwards), we can construct detailed merger trees along the
path of the main progenitor. Using the main halo at a given time, we
trace back the identified particles to the previous time step and
correlate them with the most massive halos at this time. The
progenitor is identified with that halo which contains at least $50\%$
of the particles of the parent halo. We also follow the trajectories
of their respective centers. This will assure that we follow the right
predecessor. Once a branch has been chosen along the halo's tree, we
follow it up until this branch opens again and repeat the same
procedure. Note that in this way the followed halo is not necessarily
the most massive halo at all times in the simulation, but rather most
of the time.

Figure~\ref{fig:trees} shows the merger trees for our five models. The
size of the circles is proportional to the logarithm of their
respective masses (indicated by the numbers located by the left
columns of each panel) at the respective times (indicated by the right
columns labels). The degree of accuracy of the trees depends upon the
number of identified structures at each time step. In general, our
merger trees consider all identified structures per time step, but for
the sake of clarity we have only plotted the relevant halos to
illustrate the evolution of the models.

Once the merger trees have been elaborated, one can construct the MAHs
straightforward. The red dots in Figure~\ref{fig:mah} represent the 
measured masses normalized with
respect to their respective final halo masses. The blue solid lines
represent the best fit according to the W02 model, while the dashed
black lines --- to the \cite{t04} model.  The fits were obtained by
$\chi^2$ minimization. In general, the
\citeauthor{t04} model represents a better fit to all models than the
W02 one. However, both methods fail in reproducing the early parts of
the MAHs. The deficiencies of the W02 model are emphasized in models C
and D which have a more violent history than the rest of the
models. On the other hand, the \citeauthor{t04} fits do not represent
the full MAH of such halos, see for example the large deviations at
$a\sim0.6$ for both models (the time between two major mergers). It is
clear that neither fits reproduces the MAH of a halo throughout its
entire history. At best such fits can accurately describe the MAH
within one single passive evolution phase. Having stated that, we have
used Model B, which has the longest quiescent phase, to test the
quality of the analytical fits.


\subsection{NFW analysis}
\label{sec:nfw}

For all halo models, we fit the NFW profiles as a function of time and
thus follow the cosmological evolution of their parameters (\rs\ and
\rhos).  Many factors may affect the fits of analytical profiles to
the observed (simulated) ones: the choice of binning, the merit
function, the weights assigned to the data points, the range of radii
used in the fitting \citep{t04}. We have divided each halo into
spherical shells equally spaced logarithmically until $0.6 R_{\rm
vir}$, in order to give more statistical weight to the inner regions
of the halos. We have chosen this radius since fitted profiles do not
change substantially beyond this radius. In case that a massive
subhalo is nearby (we define such halos as those with mass $> 0.3
M_{\rm vir}$) the density profiles are non-monotonic and bumps at the
data point distributions are present. We avoid such bumps by
performing fits until $min(0.6 R_{\rm vir},0.5d_{\rm h})$, where
$d_{\rm h}$ is the distance to the given subhalo. We have set the
minimum fitting radius equal to $2\epsilon$ (where $\epsilon$ is the
gravitational softening imposed in the \nbody\ force calculation) in
order to avoid two-body interactions. This minimum radius is well
within the range where a density profile for the characteristic of our
models can be considered to be ``resolved"
\citep[\eg][]{Binney:2004aa,diemand04,Reed:2005iv}.  The fitting
procedure is performed by weighted $\chi^2$, where the residual in a
given shell is normalized by its own density.

\begin{figure}[!t]
\epsscale{1.25}
\plotone{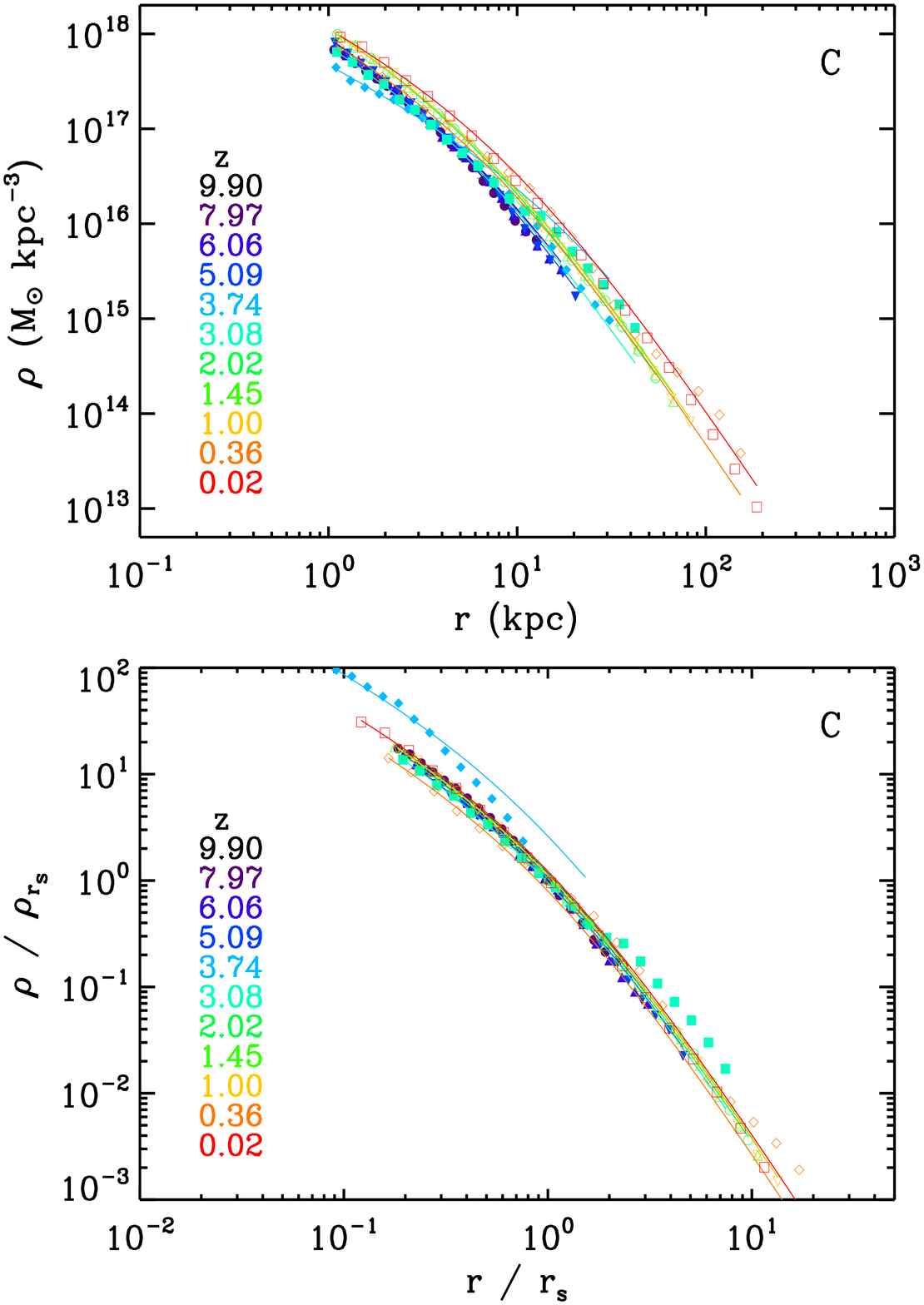}
\caption{NFW fits for model C. The symbols are the data measurements 
(symbols and colors indicate different halo epochs), while
the continuous lines represent the best NFW fit. The upper panel shows
the normal fits, while the bottom panel presents the normalized fits and
data.}
\label{fig:nfw}
\end{figure}

The upper panel of Figure~\ref{fig:nfw} shows the NFW fits for 12
different time outputs (indicated by different colors and symbols) of
model C. The bottom panel presents the same fits but normalized by \rs\
and \rhos\ respectively. Symbols indicate the data measurements while
the continuous lines the best fitted models. Most of the fits are in
good agreement with the measured data. Such fits correspond to the
steady/quiet epochs in the halo evolution. The fits that do not follow
the observed data correspond to those epochs where the halo is passing
through a violent episode (major merger).

\subsection{Virial quantities}
\label{sec:vir}

\begin{figure*}[!t]
\epsscale{1.1}
\plotone{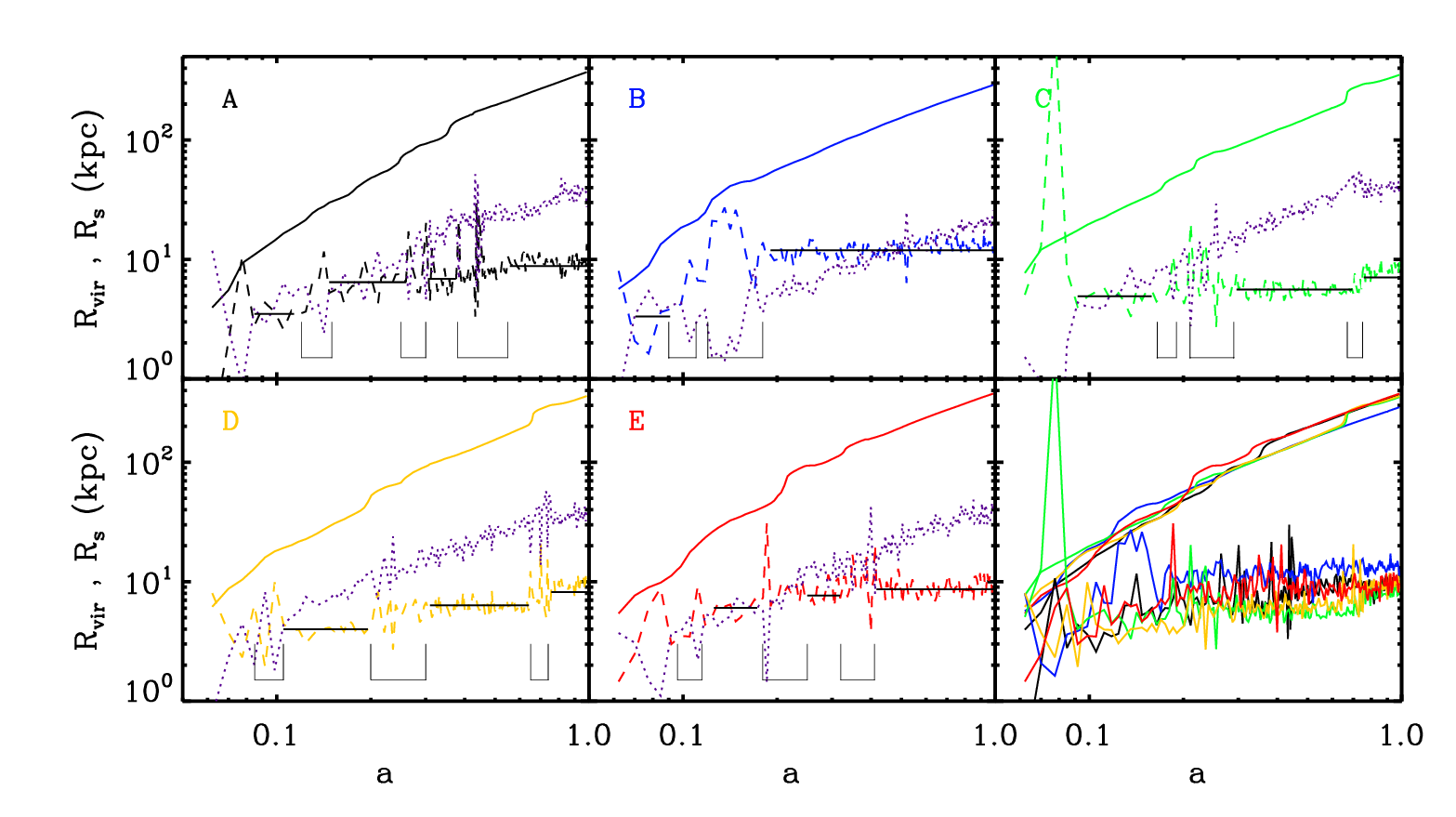}
\caption{Evolution of the virial radii \rvir\ (solid lines), 
characteristic radii \rs\ (dashed lines) and concentration parameter
\cs\ (dotted lines). Each panel refers to a different model. The
bottom right-most panel compares directly the \rvir\ and \rs\
quantities.}
\label{fig:rvir_rs}
\end{figure*}

\begin{figure*}[!t]
\epsscale{1.1}
\plotone{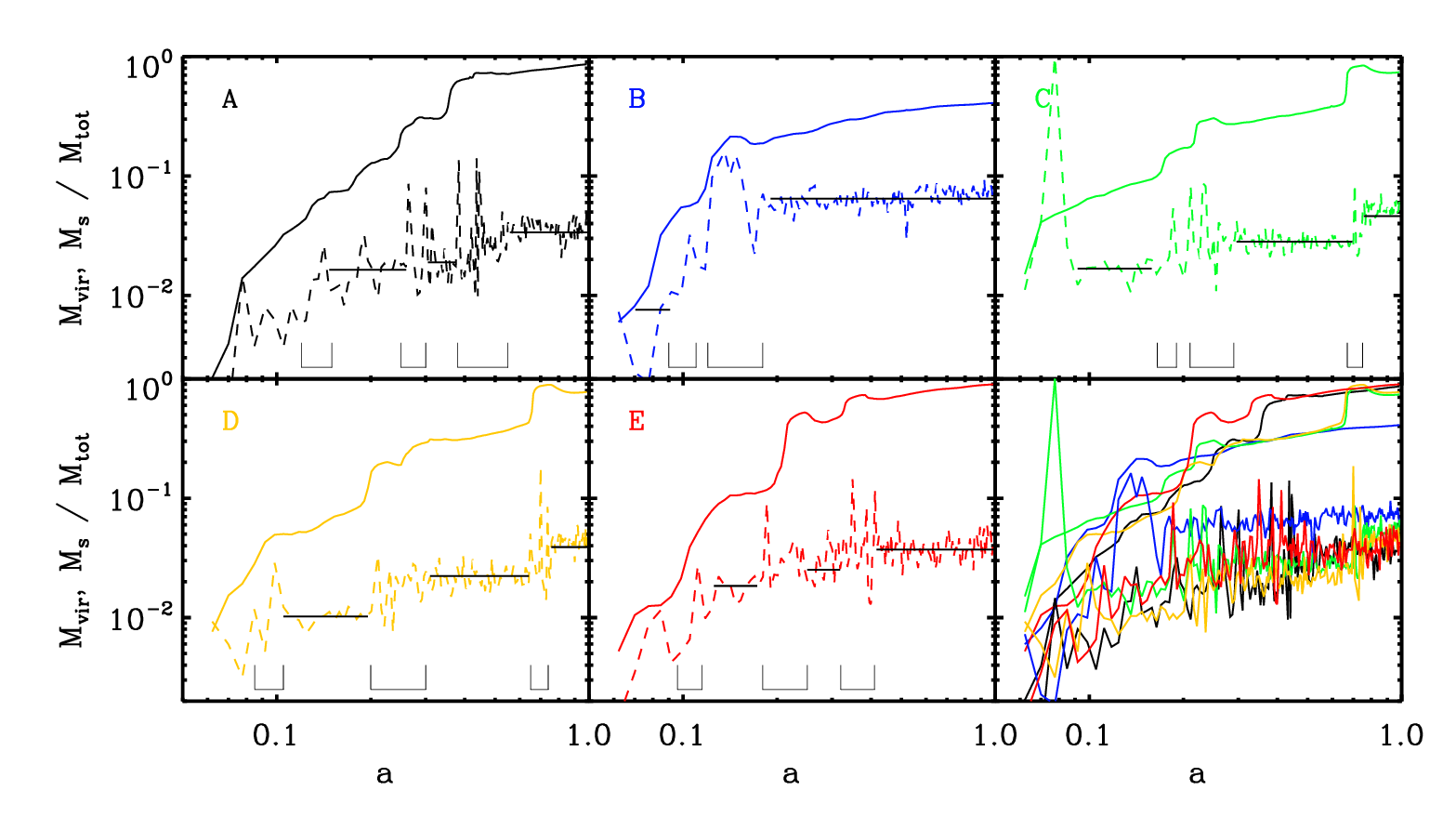}
\caption{Evolution of the virial mass \mvir\ and the characteristic 
mass \ms\ (in units of the total mass, $M_{\rm tot}$) for all models. 
The bottom right-most panel presents all
mass trajectories. Different colors correspond to different models.}
\label{fig:mass}
\end{figure*}

By following the evolution of the halo virial quantities (\rvir,
$M_{\rm vir}$), we are able to distinguish, as a function of time, the
main large scale characteristics between the different models.
Figure~\ref{fig:rvir_rs} shows the evolution of the virial radii for
the different models (solid lines). In the bottom right-most panel,
all profiles have been superimposed for a better comparison.  Note
that the virial radius grows almost linearly with \at\ during the
quiescent phases in all models. The sudden
increases correspond to the epochs where violent activity takes place
indicated by the square brackets at the bottom of each panel. The
width of the brackets shows the duration of the violent
event. Despite the fact that the various models have different
violent epochs at different times, all of them (apart from model B)
seem to converge to the same virial values once their last major
merger has taken place. This happens since the larger constraint has
been set to form a $\sim 10^{12} \msun$ halo. The violent activity is
more noticeable in the virial mass (Figure~\ref{fig:mass}, continuous
lines) where the large jumps indicate a substantial mass component
addition to the main progenitor. During the quiescent epochs, mass is
only deposited very slowly through accretion and minor mergers that
are being tidally disrupted outside \rs. This can be observed as a
very gentle increase in the mass trajectories between the major
mergers.

\subsection{Inner characteristic quantities: \rs, \ms, $c$, etc.}
\label{sec:rs}

Figure~\ref{fig:rvir_rs} also exhibits the evolution of the inner
radius \rs\ (broken lines), and the concentration parameter \cs\
(dotted lines).  Subject to a $\sim 10\%$ jitter, \rs\ is a growing
step function. In other words, \rs\ grows only at the moments when
major events take place (indicated by the bottom square brackets) and
it remains constant along the quiescent phases. The horizontal lines
which overlie the distributions are the average of \rs\ between the
major events. Note that the \rs\ distributions follow closely the mean
lines for the whole ranges. The amplitudes of the jumps are
proportional to the degree of ``violence'' of the encounters, in other
words, to the change in the kinetic energy that the system experiences
(Paper~I and Sec.~\ref{sec:energy}). The large jumps and
discontinuities in \rs\ are a clear indication of major mergers taking
place. In the bottom right-most panel, all \rs\ trajectories can be
compared more closely.  All models reach approximately the same value
after their last major merger (apart from model B), independently of
the number of major mergers and the epoch they took place. It is
interesting to see that model B has the largest \rs\ among all the
models.

\begin{figure}[!t]
\begin{center}
\epsscale{1.2}
\plotone{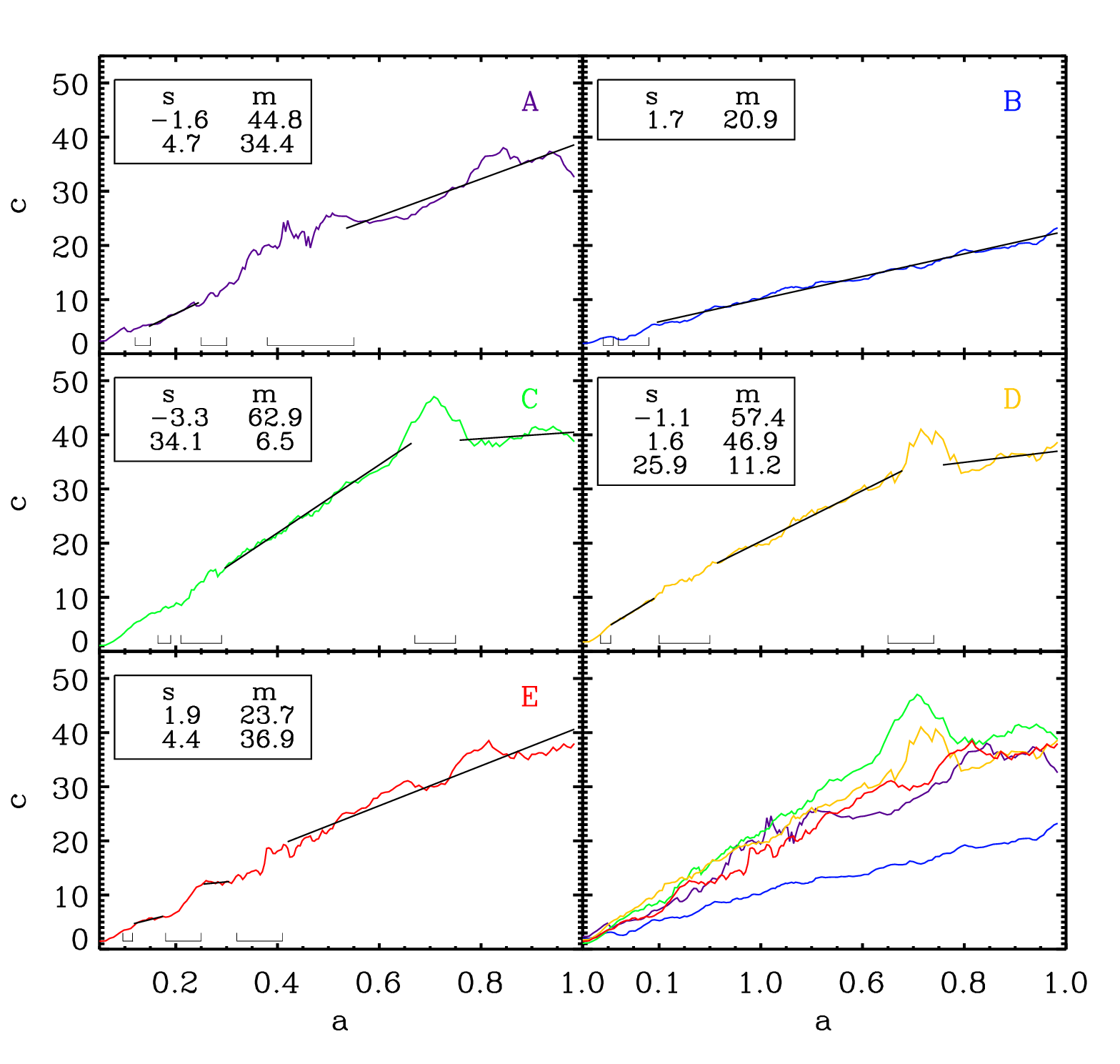}
\caption{Linear plot of the evolution of the concentration parameter, 
\cs, for the five models. The individual curves has been smoothed over
ten points. The continuous lines represent the best linear fits for 
various quiescent phases. The fitting has been avoided for short phases 
because of the associated errors. The enclosed top-left parameters 
indicate the best zero-points, $s$, and slopes, $m$.}
\label{fig:c}
\end{center}
\end{figure}

The corresponding concentration parameters \cs\ (Figure~\ref{fig:c})
display an erratic behavior at early epochs, e.g., $a \sim 0.1$.
However, starting with $a \sim 0.2$, the trajectories become more
coherent. Different model \rs\ and \rvir\ exhibit the tendency to
converge after their last major merger, while \cs\ does not (the
bottom right panel of Fig.~\ref{fig:c}).  The jitter present in \cs\
is mainly dictated by the \rs\ behavior. It has been claimed that \cs\
grows linearly with time (W02). However, since we resolve the
step-like evolution of \rs\ and the discontinuities in \rvir, \cs\
appears to grow linearly only during the quiescent phases. We show
this clearly in Fig.~\ref{fig:c}, where the black continuous lines
represent the best linear fits for the given quiescent phases. The
numbers shown within each frame represent the best fit parameters for
the slopes, $m$, and zero-points, $s$. Note that the slopes decrease
after each major event. The intermediate slopes for Models A and E are
unreliable and were not fitted because the halos do not have enough
time to relax between these major events.

Model B has been assembled earlier compared to other models, but it
possesses the smallest \cs\ at the present time.  This behavior
appears to be in contrast with the previously reported results
\citep[i.e.][W02]{bullock01sf} which claimed that the halos assembled
early, when the universe has a larger density, are more concentrated
and have a larger \cs\ parameter.  However, \rs\ and, therefore, \cs\
are related not only to the density of the universe at that time, but
also to the {\it intensity} of an event during the formation epoch of
this particular halo (Fig.~\ref{fig:c}, see also Paper~I). Since model
B has experienced the strongest event in terms of the deposited
kinetic energy $K_{\rm s}$ within \rs\ (see Fig.~\ref{fig:ke}), it has
the larger \rs\ and, therefore, the smallest \cs.

It is interesting that both models C and D seem to depart from the
general trend observed by the rest of the models at $a \sim 0.7$. This
corresponds to the epoch in which both models went through their last
major merger. After these events took place, the \cs\ trajectories
lower their amplitude and join the general trend. The growth of \cs\
seems to be nearly linear with respect to \at\, during the quiescent
phase, indicating that its behavior is dominated mainly by \rvir\ (see
also Fig.~\ref{fig:rvir_rs}).

The mass \ms\ is computed by counting particles within this radius and
hence it is a top-hat mass. The \ms\ behavior is presented in
Figure~\ref{fig:mass} (dashed lines) for all five models, and is very
similar to that of \rs --- it remains constant during the quiet phases
and grows only during the violent events. The horizontal lines
indicate (as in the case of \rs) the mean \ms\ within the given time
intervals. Similarly to Fig.~\ref{fig:rvir_rs}, the bottom brackets in
each frame indicate the time and duration of the major mergers. The
model B also has the largest \ms, as noted at the bottom right-most
panel where all \ms\ trajectories are depicted.  At the time of the
violent events associated with the core growth, the system is not in a
virial equilibrium and so the density profile is undefined causing the
erratic appearance of spikes in \rs\ and \ms\ (e.g.,
Figs.~\ref{fig:rvir_rs},~\ref{fig:mass} and ~\ref{fig:fits_all}).


\subsubsection{General trends in \rs}
\label{sec:all_models}

The \rs\ behavior detected in Paper I and in the previous section has
been only addressed from the `typical' NFW fitting procedures. As has
been pointed out in Section~\ref{sec:intro}, there is a controversy
whether the inner slope is $-1$, steeper or shallower, or it is not
universal at all. This raises the question whether the \rs\ behavior
is driven by the change in the inner slope (a mere effect of the NFW
fitting procedure) or by a real change in \rs. To answer this
question, we fit the halos using three alternative density profiles,
i.e., by applying the generalized law \citep{zhao96}:
\begin{equation}
\rho (r) = \frac{4\ \rho_{\rm s}}
          {(r/R_{\rm s})^{\gamma} (1 + (r/R_{\rm s})^{\alpha})^{\frac{\beta 
		- \gamma}{\alpha}}} \,,
\label{eq:gral_fit}
\end{equation}
where $(\alpha,\beta,\gamma) = (1,3,1)$ for a NFW, $(1.5,3,1.5)$ for a
\cite{moore99}, $(1,3,1.5)$ for a \cite{js02} and $(\alpha,3,\gamma)$
for a more general density profile.  In general, the fitting procedure
varying the parameters $\alpha, \beta$ and $\gamma$ leads to
degeneracies between the parameters \citep[see][]{kkbp01,t04}. With
this approach, it is possible to find ``excellent" fits using
parameter values far from what have been found for that kind of
halos. Therefore, in order to get physically meaningful fit estimates
one has to constraint the parameter range.

\begin{figure}[!t]
\epsscale{1.2}
\plotone{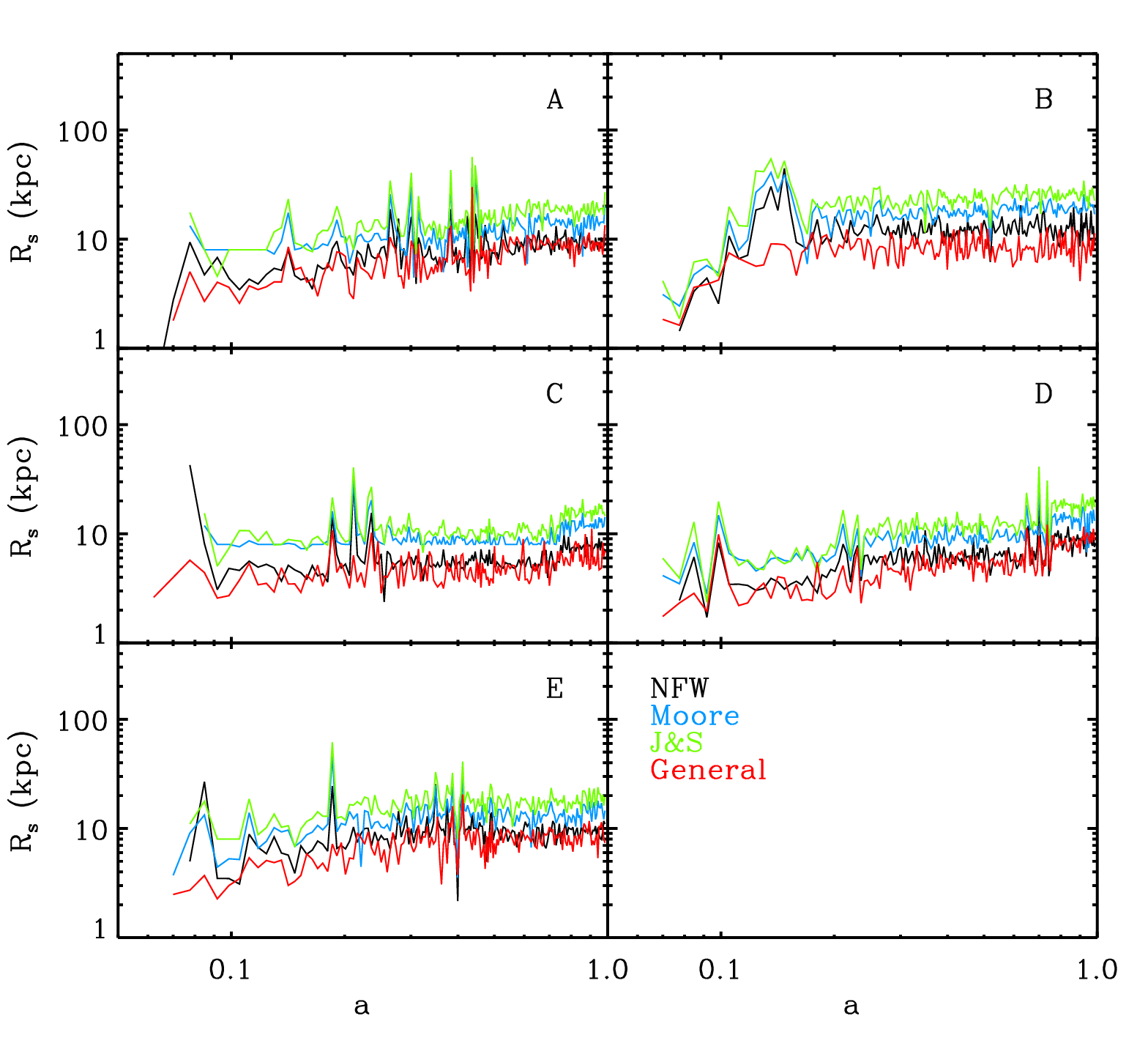}
\caption{Evolution of density profiles using various fits to our five
models.  The black lines represent the NFW profiles, the blue lines
--- the \citeauthor{moore98} profile, the green lines --- the
\citeauthor{js00} profile, while the red lines --- the general
profile of \citeauthor{zhao96}.}\
\label{fig:fits_all}
\end{figure}

Figure~\ref{fig:fits_all} shows the evolution of \rs's obtained from
four different fitting profiles indicated by different colors. Note
that all profiles show the same behavior, \rs\ remains constant during
the quiet phases and increases abruptly during the violent epochs. It
is expected that the application of different density profiles will
result in differing values of \rs, which can be correlated
\citep[\eg][]{t04}. The general fits (red lines) are very close to the
NFW fit (black lines). This effect could be due to our use of NFW
values to constrain the general fitting.

\cite{Reed:2005iv} found that the inner slope does not decrease in
time, but rather stays constant during the quiescent phase, by
applying a semi-general fitting to the halo evolution. We can apply
this observation to each of the phases and conclude that the slope
also changes only during the violent events. Combining these two
effects, one expects that the NFW \rs\ displays a similar behavior to
\rs\ obtained from other fitting procedures; Fig.~\ref{fig:fits_all}
shows clearly this point. Because we observe that \rs\ grows as a step
function even for the general fit, this indicates that not only a
change in the inner slope is involved, but rather a real
behavior/change in \rs.

\subsubsection{The characteristic density \rhos}
\label{sec:rho}

The characteristic density has the opposite behavior to \rs\ --- it is
a decreasing function of time, so their product, \rhos\rs, appears
nearly constant with time for all models and across the violent and
quiescent phases. As \rs\ also \rhos\ is affected by the $10\%$
jitter. At major mergers it drops considerably and fluctuates until it
reaches a plateau where it remains constant. At the next major merger
it drops and reaches a new, lower level plateau.

\begin{figure}[!t]
\epsscale{1.2}
\plotone{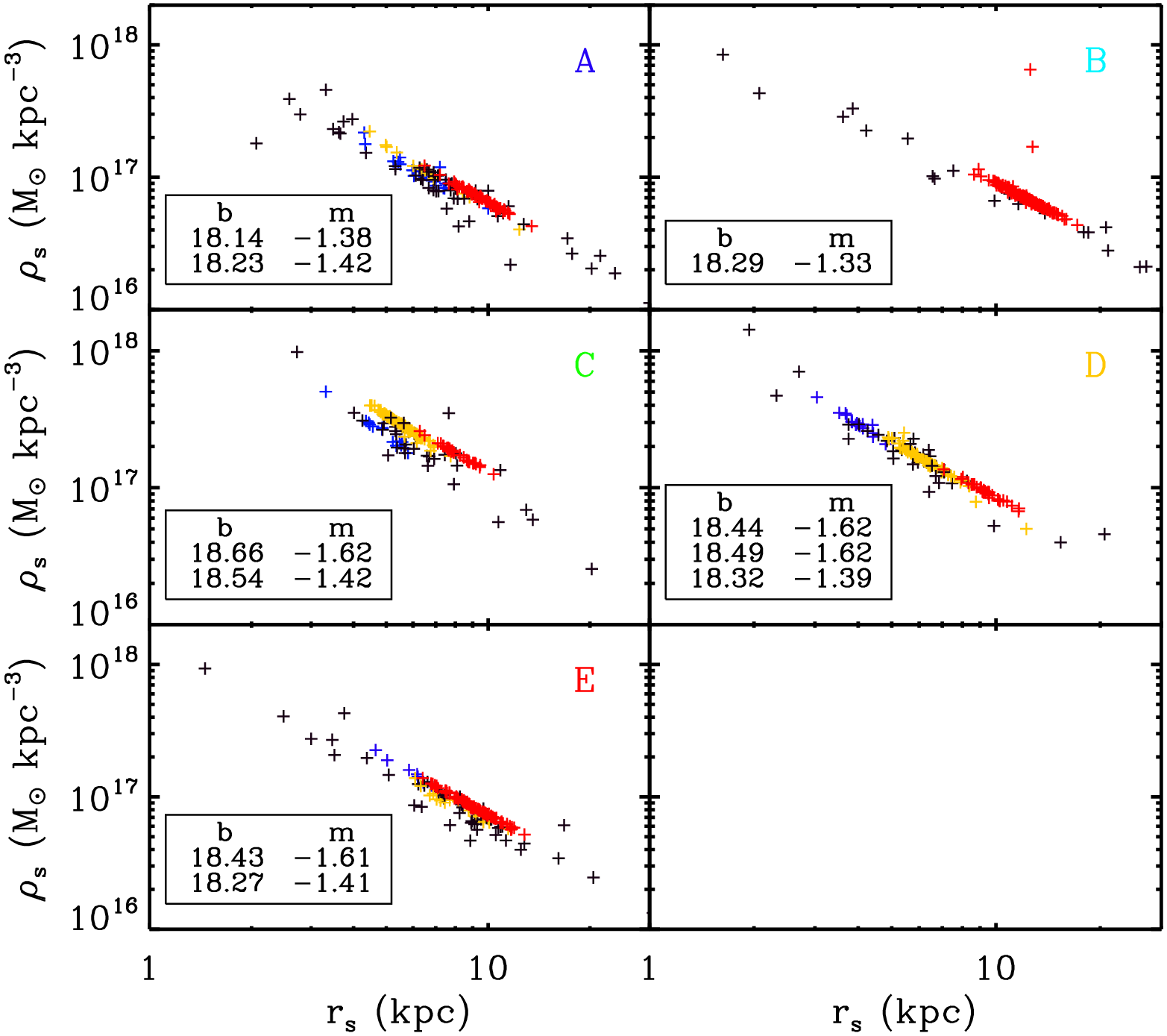}
\caption{The characteristic density \rhos\ as a function of the
characteristic radius \rs. The colors indicate the different quiescent
phases that each halo passes through. The individual slopes, $m$, and
the zero points, $s$, are specified for separate quiescent phases for
every model (see text for further details). The fitting has been
avoided for short phases because of associated errors.}
\label{fig:rhors}
\end{figure}

We have investigated the possibility of a correlation between \rs\ and
\rhos within the jitter of the quiescent phases. Naively one expects
that \rs\ and \rhos\ in the jitter will produce a scatter
diagram. However, Figure~\ref{fig:rhors} shows that a strong
correlation exists between both quantities: larger \rs\ correspond to
smaller \rhos. Moreover, a one-to-one correlation exists between \rs\
and \rhos\ across the full range of \rs.  The different colors
represent the various quiescent episodes that the halos experience
through their MAHs. The black symbols correspond to the violent
episodes and to the early assembly time of the halo. The scattered
points located above and below the main trends mark the violent
events. Model B displays a single phase since this halo formed at an
early phase and remained quiescent for most of its evolution. Models C
and D, which display three well identified quiescent phases, have
three well defined point distributions separated from each
other. Models A and E present very similar distributions. We have
performed linear fits to the logarithmic distributions for all models
of the form
\begin{equation}
 {\rm log}~\rho_{\rm s} = m\ {\rm log}~R_{\rm s} + b \,. 
\end{equation}
in two fashions. First, by including all
plotted points; second, by separating the quiescent phases and fitting
each. The upper legends at each panel indicate the slope of the
individual fits $(m)$ and their zero points $(b)$.  Note that all the
zero points lie about log~\rhos$\sim 18.2$. The slope averaged over
{\it all} models is $m =-1.37 \pm 0.03$. Even the slope constructed
from different epochs (given by different colors in
Fig.~\ref{fig:rhors}) shows very little scatter around the average
value. {\it This value appears to be independent of particular halos
and of their evolution.}  Furthermore, the points corresponding to the
later time (i.e., red color) are consistently located in the tail of
their respective distributions.  This color order is a direct
consequence of \rs\ and \rhos\ monotonic evolution discussed
earlier. The measured jitter ($10\%$ or so) in both \rs\ and \rhos\
can be recognized now as the individual trends (colors) for each
model. This implies that a clear order exists in the halo
variations. Specifically, we find that \rhos$\sim $\rs$^{-1.59\pm
0.15}$ during the jitter at early quiescent, i.e., $a < 0.5$, epochs
in all models, while \rhos$\sim $\rs$^{-1.39\pm 0.04}$ for later
times. When translated to \ms---\rs\ correlation, this amounts to
\ms$\sim $\rs$^{1.41}$ and \ms$\sim $\rs$^{1.61}$, respectively. The
simplest interpretation of these correlations is that they are driven
by density fluctuations which originate at radii between the cusp
(i.e., slope = --1) and \rs\ (i.e., slope = -2).
 
Lastly, the shifts between different colors in Fig.~\ref{fig:rhors}
come from violent events which deposit kinetic energy and mass within
\rs\ and separate the quiescent evolution (see section 5.5).  Our
results are in agreement with those reported by \cite{z03b}, who found
that \ms $\sim $\rs$^{1.44}$ holds in the ``slow accretion phase"
(gentle accretion), and \ms$\sim $\rs$^{1.92}$ in the ``rapid
accretion phase."  The difference results from our ability to separate
the violent and quiescent phases --- we have excluded the black
symbols in Fig.~\ref{fig:rhors} from the fitting, because they
correspond either to the very early assembly or later violent
times. This reduces the scatter in our fits below that of Zhao et al.

\subsection{Kinetic and Potential energy and the virial ratio}
\label{sec:energy}

We have computed the internal kinetic energy $K$ of the main halo
along its merger tree. For each time output, we have computed the
kinetic energy, $K_{\rm r} = \sum_{i=1}^{n_{\rm r}} \frac{1}{2}m_{\rm
i} \sigma_{\rm i}^2$, where the summation goes for all objects
enclosed within a radius $r$, $n_{\rm r}$, and $\sigma_{\rm i}$
represents the velocity dispersion of a given object $i$ of mass
$m_{\rm i}$.

\begin{figure}[!t]
\epsscale{1.2}
\plotone{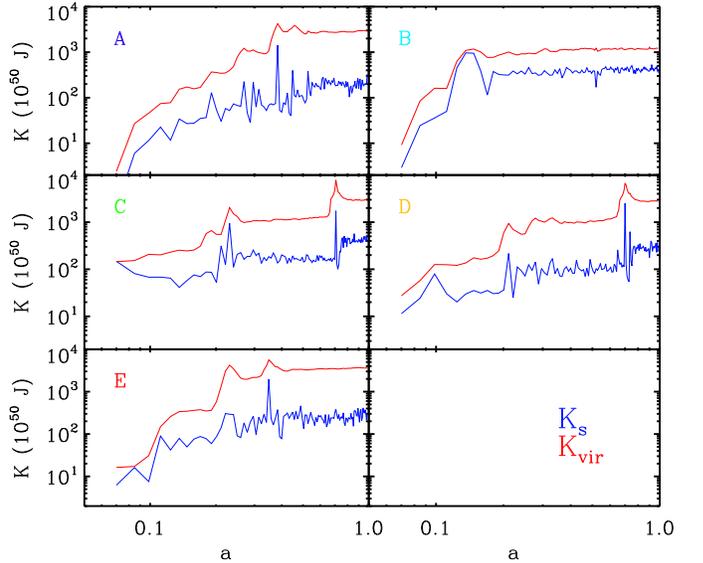}
\caption{Kinetic energy for all models as a function of time. The blue
 distributions represent the characteristic kinetic energy, while the
 red ones the virial distributions. Labels are the same as in the
 previous figures.}
\label{fig:ke}
\end{figure}

Figure~\ref{fig:ke} shows the evolution of the internal kinetic energy
computed within the radii \rs\ (blue lines) and \rvir\ (red lines). As
in the case of \rs, \ks\ also increases as a step function, \ie\ only
at the major events and it remains constant during the quiescent
phases. On the other hand, \kvir\ grows in two different ways, it
grows suddenly at the violent phases as \ks\ does, while during the
quiescent phases it grows in a very gentle way.

From Fig.~\ref{fig:ke}, one can realize that \ks\ of model B has the
largest value of all models while its \kvir\ is the smallest from all
models. This is a mere reflection of their mass accretion
histories. Furthermore, the growth of model B energy curve is the most
abrupt of all models, occurring at an early redshift ($z\sim 9$). This
halo went through a very violent and rapid formation epoch at this
early $z$. This violent formation event resulted in the largest \rs\
in our sample, although it is the less massive of the five models.

\begin{figure}[!t]
\epsscale{1.2}
\plotone{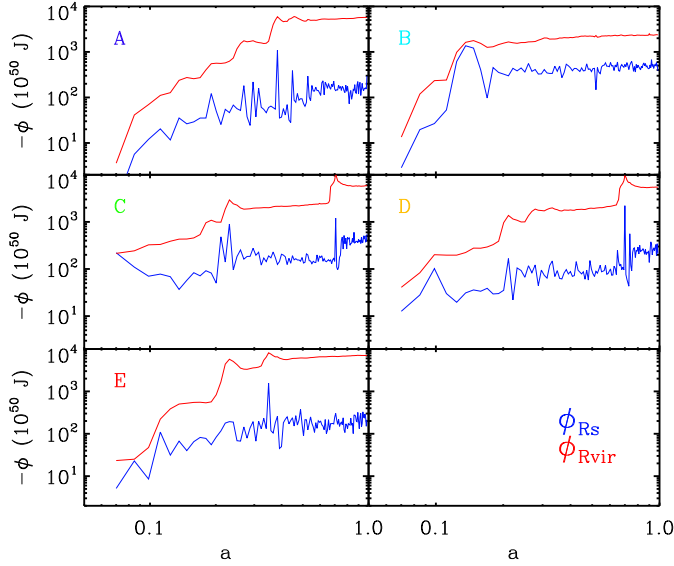}
\caption{Potential energy for all models as a function of time.
Labels and colors are the same as in previous figures.}
\label{fig:pe}
\end{figure}

The potential well of the halos is also constructed from their
respective MAHs. \cite{z03b} found in their simulations that the
potential well is mainly build up in the so-called fast accretion
phase, while remaining almost constant during the slow accretion
phase. Figure~\ref{fig:pe} shows the evolution of the potential energy
($\phi$) for our set of models. This $\phi$ is computed by including
all particles within a given radius, namely \rs\ and \rvir. Since
there is no mass accretion within \rs\ in the quiescent phases, it is
expected, therefore, that the potential \phis\ remains constant during
such phases and only deepens at the major events when mass is accreted
within \rs. In the case of the potential within \rvir, where the halo
growth is due to violent and gentle mass accretion, \phiv\ uncovers
these two phases as well. It sinks substantially at the major events
and it deepens in a very gentle way during the quiescent phases. This
general behavior indicates that the potential wells are only formed
and substantially modified at the violent events. The main difference
with the \cite{z03b} analysis is that they only recognize a single
early fast accretion period, while in our models the later quiescent
phase is intermitted by the violent events.

The way energy is acquired by the halos and the details of the
formation of the respective potential wells reinforces our
view of the halo growth process. Energy is deposited into the halo via
gentle and violent mass accretion. The latter affects the whole halo
since both \rs\ and \rvir\ grow during such events, while the former
affects only the exterior part of the halo. Gentle accretion and minor
mergers deposit their mass and energy within this outer
region. Energetic events such as those produced by the major mergers
are able to reach the halo core, since they are more strongly
gravitationally bound. Part of their mass and energy are deposited
inside the halo core.

\begin{figure}[!t]
\epsscale{1.2}
\plotone{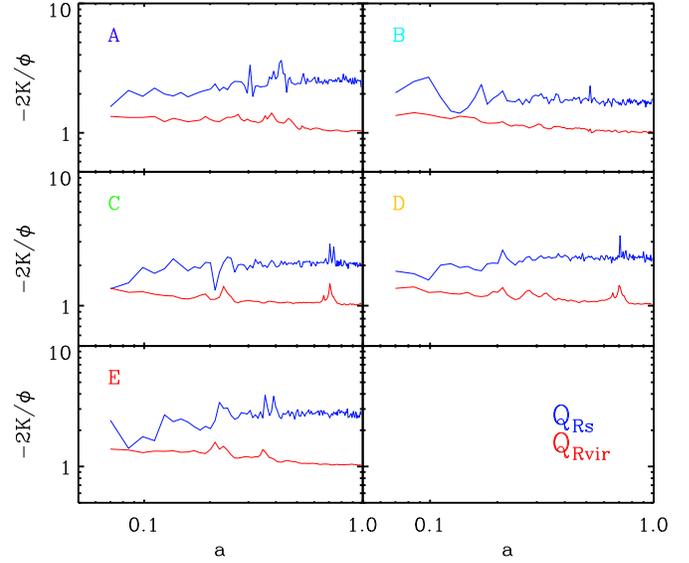}
\caption{Virial ratios for all models as a function of time. Labels 
and colors are the same as in previous figures.}
\label{fig:q}
\end{figure}

The interplay between the kinetic and potential energies within a
given halo can be better understood in terms of the virial ratio
defined as $Q_{\rm r} = - 2 K_{\rm r}/\phi_{\rm r}$, where the
subindex $r$ denotes the radius at which this quantity is computed. At
the virial radius, $Q_{\rm r}$ should be close to unity for systems in
virial equilibrium and nearly isolated. Because virialized
structures are bordered by infalling and outgoing material,
the convergence of $Q_{\rm r}$ to unity at \rvir\ should
not be fully expected \citep[see \eg][]{mmb03,Shaw:2005ua}. However, 
since our models are basically constructed in isolation,
after their respective last major merger, \qvir$\rightarrow
1$. Indeed, this can be observed in Figure~\ref{fig:q} where the solid
lines represent the \qvir\ quantities for each model. The pronounced
spikes along each track represent the time of the major events and the
considerable departure from the virial equilibrium. On the contrary,
\qs\ is not expected to be close to unity, since this region experiences
``pressure'' from overlying halo. Note that
during the quiescent phases \qs\ remains constant and, as in the case
of \rs, jumps to another stationary trajectory when the major events
take place.

\subsection{Shapes}
\label{sec:shapes}

We have analyzed the shapes of our halos as a function of time by
using the inertia tensor, for \rs\ and \rvir\ respectively, using the
unweighted moment of inertia defined as $I_{\rm ij} =
\sum_{n=1}^{n_{\rm r}} x_{\rm i,n} x_{\rm j,n}$, where $x_{\rm i},
x_{\rm j}$ represent the $i,j$ Cartesian components of the position of
a particle $n$ and the summation extends over all particles enclosed
within the given radius $r$ ($= $ \rs, \rvir). \cite{allgood06} showed
that the weighted and unweighted moments of inertia used to compute
halo shapes differ only by $10\%$.  Our aim here is to compute the
halo shape evolution at various radii as a function of their merging
history. Therefore, we divide the halos into spherical shells. An
alternative approach is to compute the shapes within isosurfaces of
density or potential \citep{ingo06}.  Since so far we used quantities
defined within the \rs\ and \rvir\ spherical distributions, we choose
the former approach for consistency.

\begin{figure}[!t]
\begin{center}
\epsscale{1.2}
\plotone{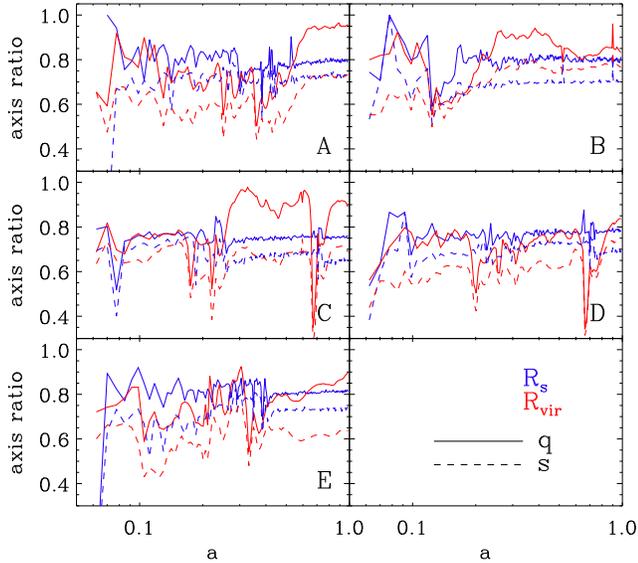}
\caption{Axes ratios for our five models. For each model we present
$s~ (\equiv c/a,$ dashed lines) and $q~ (\equiv b/a,$ solid lines)
ratios for \rs\ (blue lines) and \rvir\ (red lines).}
\label{fig:axis_rat}
\end{center}
\end{figure}

We compute the eigenvectors and eigenvalues of the moment of inertia
matrix and define the axes of the ellipsoid as $a,b,c =
{\sqrt(\lambda_{\rm a},\lambda_{\rm b},\lambda_{\rm c})}$, where
$\lambda_{\rm i}$ corresponds to the $i$th eigenvalue, and they are
arranged in the descendent order (\ie\ $\lambda_{\rm a} > \lambda_{\rm
b} > \lambda_{\rm c}$). These axes are normally described in terms of
the ratios $s \equiv c/a$, $q \equiv b/a$ and $p \equiv c/b$.

Figure~\ref{fig:axis_rat} displays the shape evolution of \rs\ and
\rvir\ indicated by the red and blue colors respectively. The dashed
lines correspond to the $s$ parameter of both radii, while the solid
lines --- to the $q$ number. Clearly, all models are prolate within
\rs. This result is in agreement with the analysis of
\citet{allgood06} at small radii, and of \citet{ingo06}.  Within
\rs\ the halo shapes are unperturbed and only change at the major
mergers.  On the other hand, the \rvir\ shapes change during the
quiescent epochs as well, mainly because gentle accretion is not
isotropic.  The $s$ and $q$ at \rvir\ indicate that there is a
tendency for the halos to become less triaxial with time.


\subsection{Angular momentum considerations}

One of the drawbacks of the present simulations is the lack of the
external tidal field presence in our simulations. Although our
simulations have accounted for the tidal field within the original box
of $8 ~\hmpc$, this volume does not suffice to include the large scale
tidal contributions. Nevertheless, this affects only the final phases
of the formation of the halo. At earlier times, the main halo is
surrounded by other halos of about the same mass that torque it.
Therefore, the measured angular momentum will be mainly produced by
the direct interactions of the halo with those substructures created
with the constrained random field. Any differences between the models
will be attributed to their different merging histories.

\begin{figure}[!t]
\begin{center}
\epsscale{1.2}
\plotone{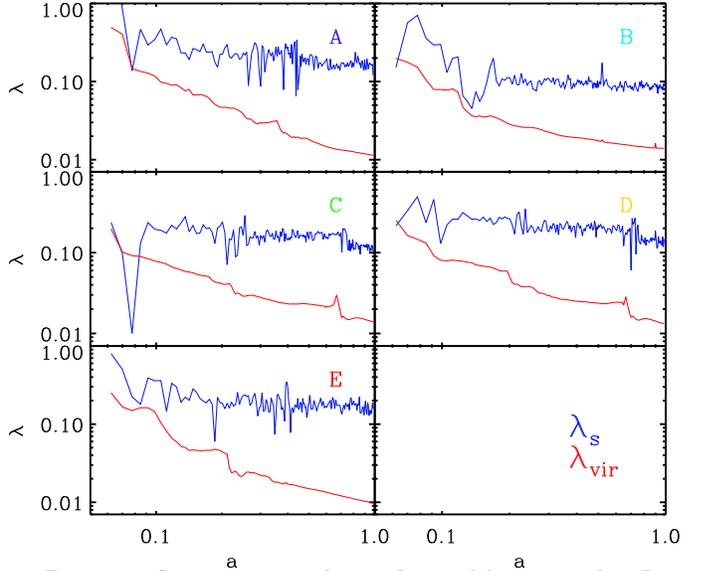}
\caption{Spin parameters for our five models computed at \rvir\ and \rs.}
\label{fig:ang_mom}
\end{center}
\end{figure}

The angular momentum can be expressed in term of the dimensionless
spin parameter \citep[\eg][]{peebles69} $\lambda \equiv J
|E|^{1/2}/(G\, M^{5/2})$, where $J,~E$ and $M$ are the total angular
momentum, energy and mass of the system, and $G$ is the gravitational
constant. The value of the spin parameter roughly corresponds to the
ratio of the angular momentum of a halo to that needed for rotational
support \cite[see \eg][]{p93}. Typical values of the spin parameter of
individual halos from \nbody\ simulations are in the range
$[0.02,0.11]$
\citep{barnes87,ryden88,w92,sb95,cl96,gardner01,bullock01ck}. The spin
parameters in \nbody\ simulations are independent of the cosmological
model \citep{barnes87,w92,lk99,gardner01}, halo environment or halo
mass \citep{lk99}. The only correlation that has been found is with
the time of the last major merger: halos that had a recent major
merger have larger spin parameter \citep{gardner01}.

\cite{bullock01ck} proposed an alternative and more practical way to
compute the spin parameter $\lambda' \equiv J/(\sqrt{2}\, M\, V\, R)$,
where $J$ is the angular momentum within a spherical volume of radius
$R$ with mass $M$ and $V$ is the halo circular velocity at radius $R$,
$V^2 = GM/R$. This definition reduces to the standard one when
measured at the virial radius of a truncated singular isothermal halo
\citep{bullock01ck}. In $\lambda'$ definition, $J$ is defined as ${\bf
J} = \sum_{i=1}^{n_{\rm r}} {\bf r}_{\rm i} \times m_{\rm i} {\bf
v}_{\rm i}$.  We have adopted the \citeauthor{bullock01ck} definition,
hereafter $\lambda$. Figure~\ref{fig:ang_mom} shows the evolution of
the spin parameter computed at the two usual radii, \rvir\ and \rs,
\ls\ and \lvir (blue and red lines respectively).  The sudden changes
in \lvir\ are clearly correlated with the major events involving each
halo. \lvir\ exhibits the tendency to decline during the quiescent
epochs \citep[see also][]{v02}. This is a consequence of fast increase
in \rvir$\sim a$ during these epochs, which makes the denominator in
$\lambda$ to grow faster than $J$ for any reasonable scenarios of the
halo growth.  Models A and E have the lowest virial spin parameter
because most of the accreting material has already joined them. Model
B has the largest \lvir\ by the end of the simulations --- it has the
smallest main halo and a substantial amount of material is locked in
its companion and around it.
 
The behavior of \ls\ differs somewhat from that of \lvir. It confirms
that the halo core grows only via accretion of major mergers and minor
mergers do not play a role. Under these conditions, one expects that
$\lambda$ will be conserved.  The blue lines of Fig.~\ref{fig:ang_mom}
shows precisely this type of behavior. The spin parameter decreases
only at the major events and remains constant during the quiescent
phases in most of the models which exhibit the same \ls\
value. However, Model B has the lowest \ls, which is due to a smaller
number of major mergers it has experienced.

This analysis shows that although the models miss large scale torques,
this does not represent a considerable problem for their
evolution. While the models have been constrained to the same initial
sizes and masses, no constraint has been imposed on the spin
parameter. Because the number of major mergers is the same in all
models except the Model~B, the final spin parameter differs little
between them.


\section{Discussion and Conclusions}
\label{sec:disc}

We have investigated the effect of a different assembly history on the
final structure of the DM halos. We have employed the Constrained
Realization method of random Gaussian fields in order to create the
initial conditions for a $10^{12}\hmsun$ halo, which has been evolved
subsequently by means of the \nbody\ simulations. Five different
variants of the same final halo have been simulated and analyzed using
a sufficient mass resolution to identify the halos and subhalos at
early redshifts ($z > 10$), and a sufficient time sampling in order to
closely follow their dynamical evolution. The model evolution has been
quantified in terms of parameters which characterize the NFW as well
as other, non-NFW density distributions.

The evolution of a given halo can be characterized by a number of
quiescent phases of a slow evolution intermitted by violent episodes
of major mergers. We find that the inner halo is in a state of a
dynamical equilibrium during the former phases, and its density is
well approximated by an NFW profile. Furthermore, the NFW
characteristic radius \rs\ appears to be the best gauge of the
dynamical evolution of a halo. It remains constant in the quiescent
phases and it grows discontinuously during the violent episodes. Other
variables that characterize the {\it inner} structure, \ie\ within
\rs, behave in a similar way in their growth or decline, e.g., \rhos,
\ms, \ls, etc.  Between the major mergers, the halos evolve very
passively.  This means that a gentle accretion does not influence the
behavior of the characteristic quantities. The minor mergers only
contribute to the gentle growth of the external halo and to a decrease
of the spin parameter. We note that the product \rhos\rs\ is nearly
constant in time showing only a very slight decline. The significance
of this will be discussed elsewhere.
 
On the other hand, the virial parameters, e.g., \rvir, \mvir, \lvir,
etc., exhibit a monotonic growth or decline during the quiescent
phases, in accordance with the simple analytical relation \citep[\eg
W02,][]{bullock01sf}. During this time, the virial radius \rvir\ shows
a nearly linear growth with the expansion parameter $a$, and the mass
accretion history, namely \mvir, is fairly well described by the
fitting formula of \citet{w02} and \citet{t04} which exhibits an
overall slowdown for the later times. However, these simple relations
hold in the quiescent phases only, and, in general, cannot be
extrapolate from one phase to the other. During the violent episodes,
the virial parameters change discontinuously. Therefore, one cannot
fit the entire evolution of a halo, consisting of a a series of
violent and quiescent phases, with a single analytical expression. The
occurrence of a few generations of violent phases is typical for the
CDM cosmogonies rather than an anomaly and any theory which aims at
modeling the evolution of halos should account for that.

The above discussion implies that the concentration parameter $c$
closely mimics the behavior of \rvir\ during each quiescent phase by
showing a nearly linear growth with $a$.  This growth, however, slows
down with each consecutive quiescent phase. The value of $c$ at the
onset of each slow phase appears to depend both on the structural
change in the halo that underwent the violent merger (i.e., change in
\rs) and on the degree of violence of that merger.
 
We note that the trends observed here do not pertain only to a NFW fit
but remain when various fitting formulae for the density profiles have
been applied. The actual numerical values of the structural parameters
(\ie\ \rs, \rvir, etc.) have been found to depend on the exact fitting
procedure, but the general trends did not change. Therefore, our
conclusions reflect the robust characteristics of the halo evolution.

The original goal of this work was to impose different constraints on
the mass distribution within the computation box with the same total
mass.  Whether this would lead to a diverging evolution was one of the
main objectives of the current project.  One of the aspects was to
obtain a different number of major mergers within the box. The crucial
question was to what extent the final halo properties depends upon its
assembly history.  Our results have shown that the imposed constraints
on the mass distribution had only a partial effect on the initial
conditions because of the random component.  As a result, the number
of major mergers after $z\sim 10$ was the same (i.e., three mergers)
in all the models, except in model B (two overlapping mergers or
one). Hence the initial conditions did not possess a sufficient
scatter to form the final halos characterized with wide range of
internal structure. Nevertheless, clear trends emerge.

First, one is unable to reconstruct the evolution of the halo without
accounting for the amplitudes of the discontinuities triggered by the
major mergers. Their overall effect is non-additive, in the sense that
larger fractional energy inputs, $\Delta K_{\rm s}/K_{\rm s}$, during
the mergers result in the non-linear growth of \rs\ and other
structural parameters as well.

In order to quantify the violent phases of halo evolution, we
introduce the `violence' parameter, $Q_{\rm vio}\equiv \Delta K_{\rm
s}/K_{\rm s}$, defined in terms of the fractional kinetic energy
deposited within \rs\ in a major merger. Assuming virial equilibrium
this parameter measures the deepening of the gravitational potential
in a violent event.  We denote the change in the characteristic
parameters across the major merger event by $\Delta$\rs/\rs $\equiv
(R_{\rm s2} - R_{\rm s1})/R_{\rm s1}$, where subscripts `1' and `2'
refer to `before' and `after' the event.  Fig.~2
of Paper~I shows that the fractional change of \rs\ in terms of the
fractional change of $K_s$ are approximately related by
\begin{equation}
Q_{\rm vio}\equiv \bigg(\frac{\Delta K_{\rm s}}{K_{\rm s}}\bigg) \propto
      \bigg(\frac{\Delta R_{\rm s}}{R_{\rm s}}\bigg)^{1/2} \,,
\end{equation}
where typically $0 < Q_{\rm vio} \ltorder 1$, but it also can be
somewhat larger than unity in principle. Thus the change in \rs\
varies in a non-linear fashion with $Q_{\rm vio}$.  Therefore, the
evolution of \rs\ depends both on the number of the violent phases and
the magnitude of each one. This implies that the evolution of \rs\
cannot be reconstructed from integral quantities but it depends on its
detailed merging history.

Second, we compare our models at four benchmark redshifts, namely
$z=0,1,2$ and 3. For this purpose, we only use the primary halos with
the same virial parameters, i.e., of the same virial mass.
Figure~\ref{fig:mass} shows that models B, C and D possess similar
mass halos at $z\sim 1-3$. However, the structural differences between
these halos at higher redshifts are more substantial than at
$z=0$. For example, The NFW scaling parameters \rs\ exhibit much
larger scatter at earlier times. They converge towards $z=0$, except
for model B. The latter has the largest \rs\ despite going through an
early prolonger major merger --- an indication that the energetics of
mergers characterized by $Q_{\rm vio}$ can be as important to the
halo's structural evolution as the number of mergers
(Fig.~\ref{fig:rvir_rs}). At $z\sim 1$ models B, C and D observe the
relation $R{^{\rm C}_{\rm s}} \sim R{^{\rm D}_{\rm s}} \sim 0.5
R{^{\rm B}_{\rm s}}$. This is also the case of models D and E at $z
\sim 5.$, for which $R{^{\rm D}_{\rm s}} \sim 0.6 R{^{\rm E}_{\rm
s}}$. These examples show how the merging histories of otherwise
similar objects can affect their NFW structure, i.e., the \rs\ related
quantities.

Third, the value of the concentration parameter \cs\ has been
demonstrated to depend not only on the formation time of the halo but
also on the degree of violence in its merger events, $Q_{\rm
vio}$. Hence, the halos forget the initial conditions to some
extent. This trend is expected to become more obvious with the
increase of the computational box, and, hence with the associated
increase in the number of major mergers that the halo goes through.

Lastly, we comment on the effect of the environment on the growth of
the halos. The virial parameters in our models grow nearly linearly
(on linear plots!) during the successive quiescent phases. A closer
look on Figs.~\ref{fig:rvir_rs} and \ref{fig:c} reveals that this
growth tends to flatten at later times in all models except model B
which stays linear since its early merger. This latter model is the
only one where the largest constraint is still collapsing at
$z=0$. Hence halos which are located in rich environments are expected
to continue growing even at present time while for those which are
located in voids, the growth will saturate.

The emerging picture from the evolution of characteristic masses and
radii and of virial masses and radii confirms that only the major
mergers are able to penetrate the core, as has been argued by
\citet{sw98}.  The minor structures seem to be stripped completely in
the external halo before they could reach the inner core, and so the
mass of the external halo region grows during the minor mergers and
the gentle mass accretion. This relatively straightforward picture
will be complicated if baryons are included. Dissipative processes
associated with them will lead to the formation of a substantially
more bound systems, either purely baryons or mixed with the DM. Such
systems will be able to survive during very un-equal (minor) mergers
and contribute to the evolution of the halo core.

The main limitation of the present simulations is the small size of
the computational box. In none of the models the halo is embedded
within the typical cosmic web and it evolves rather as an increasingly
isolated halo. The size of the box limits the effect of the tidal
interactions which induces the halo's angular momentum. Nevertheless,
the resulting spin parameter of the halo in all simulations, $\lambda
\approx 0.01$, is still within the scatter measured in cosmological
simulations.  One can conjecture that to the extent that the large
scale tidal field affects the internal structure of a halo, the
increase in the size of the computational box might lead to a stronger
dependence of the halo structure on its merging history and hence on
the imposed constraints.

The five models considered here provide five different possibilities
for the merging history leading to the formation of essentially the
same halo. Thus, only one main halo is considered here, based on one
given realization of the primordial perturbation field. More
realizations are needed to construct an ensemble of halos so as to
calculate the evolution of the averaged quantities. It follows that
actual numerical values of quantities of interest, such as the
concentration parameter, \rs, and so, cannot be considered as
representative. Yet, the different evolutionary trends pointed out
here provide an accurate description of the evolution of DM halos in
the CDM-like cosmologies.


\acknowledgments 

We thank the referee, Adi Nusser, for his comments which improved the
quality of this paper.  We are grateful to D.~Eisenstein and P.~Hut
for making their group finder algorithm (HOP) publicly available. This
research has been supported by ISF-143/02 and the Sheinborn Foundation
(to YH), by NASA/LTSA 5-13063, NASA/ATP-06GJ35G, HST/AR-10976 (to
IS), and by NSF/AST 02-06251 (to CH and IS). ERD acknowledges the
Golda Meier Fellowship at the Hebrew University.


\appendix

\section{Constrained Realizations Formalism}

Consider the primordial cosmological density field $\delta(\br)$ on
which a set of linear constraints are to be imposed. The constraints
consist of a set of positions, mass scales and the value of the
$\delta$ field at the specified locations, appropriately smoothed on
the mass scales, namely $\big\{ \br_\alpha, \ M_\alpha, \
\Delta_\alpha \big\} _{\alpha=1,...,M}$, where $\br_\alpha$ is the
location, $M_\alpha$ the smoothing mass scale and $\Delta_\alpha$ is
the value of the smoothed density field of the $\alpha$-th
constraint. The smoothed field is define by a convolution with a
Gaussian kernel,
\begin{equation}
\label{eq:apndx1}
\Delta_R(\br) = {1\over (2 \pi)^3} \int\rd^3 k\  \delta_\bk  \exp
\biggl[- \imath \br \cdot \bk - {k^2 R^2\over 2} \biggr],
\end{equation}
where $\delta_\bk$ is the Fourier transform of the primordial
perturbation field, $\delta$, and $R$ is the linear smoothing scale.
For a Gaussian smoothing, $R$ is related to the smoothing mass scale
by
\begin{equation}
\label{eq:apndx2}
R=0.64 \biggl[ {M \over (4 \pi / 3) \rho_{\rm crit} \Omega_{\rm m}} \biggr]^{1/3} \,,
\end{equation}
where $\rho_{\rm crit}$ is the critical cosmological density and
$\Omega_{\rm m}$ is the matter density parameter (Bardeen \etal\
1986).  The constraint on the mass scale $M$ is referred to as on the
co-moving linear scale $R$.

\citet{bert87} proposed that a constrained field can be composed from
the sum of a mean field (fixed by the values and forms of the
constraints) and a residual field. This last field adds the random
component to the mean field. A constrained realization of the
unsmoothed $\delta$ field has been obtained by \cite{hr91}:
\begin{equation}
\label{eq:apndx3}
\delta^{\rm CR}(\br) =\tilde\delta(\br) + \Big< \delta(\br)\ \Delta_{\rm R_\alpha} \Big>\
                       \Big< \Delta_{\rm R_\alpha}\ \Delta_{\rm R_\beta} \Big>^{-1}
                       (\Delta_{\rm R_\beta} - \tilde\Delta_{\rm R_\beta}) \,.
\end{equation}
Here, $ \tilde\delta(\br)$ is a random realization of the $\delta$
field and $\tilde\Delta_{\rm R_\alpha}$ is a mock constraint obtained
from the random realization,
\begin{equation}
\label{eq:apndx4}
\tilde\Delta_{\rm R}(\br_\alpha) = {1\over (2 \pi)^3} \int\rd^3 k\  \tilde
                   \delta_\bk \exp\biggl[- \imath \br_\alpha \cdot \bk - 
		     {k^2 R{^2_\alpha}\over 2} \biggr].
\end{equation}
Namely, $\tilde\Delta_{\rm R}(\br_\alpha)$ is the value of the random
field at the position $\br_\alpha$ after applying a Gaussian kernel
with a smoothing scale $R_\alpha$. The angular brackets denote an
ensemble average used to calculate the autocorrelation matrix of the
constraints and the cross correlation matrix of the constraints and
the underlying $\delta$ field.  The constraints auto-correlation
matrix is given by $ \Big< \Delta_{\rm R_\alpha}\ \Delta_{\rm R_\beta}
\Big> =\xi_{\alpha\beta} (\vert \br_\alpha -\br_\beta \vert), $ where
the constraints auto-correlation function is
\begin{equation}
\label{eq:apndx6}
\xi_{\alpha\beta}(\br) = {1\over (2 \pi)^3} \int\rd^3 k\  P(k) \exp\biggl[- 
      \imath \br \cdot \bk - {k^2 (R{^2_\alpha} + R{^2_\beta}) \over 2} \biggr],
\end{equation}
where $P(k)$ is the power spectrum.  Similarly, the constraints field
cross-correlation matrix is given by $ \Big< \delta(\br)\ \Delta_{\rm
R_\alpha} \Big> = \xi_\alpha(\vert \br - \br_\alpha \vert), $
where the constraints auto-correlation function is given by:
\begin{equation}
\label{eq:apndx8}
\xi_\alpha(\br) = {1\over (2 \pi)^3} \int\rd^3 k\  P(k) \exp\biggl[-  
\imath \br \cdot \bk - {k^2 R{^2_\alpha}\over 2} \biggr].
\end{equation}

One should note that a constrained field is ``effectively''
constrained not only at the place where the constraints have been
imposed, but also throughout the whole region within their respective
correlation length. Regions located beyond that length obey the random
part of the procedure. In principle, this can add other kind of
structures that can affect the dynamical evolution of the imposed
constraints.

\begin{figure}[!t]
\epsscale{1.0}
\plottwo{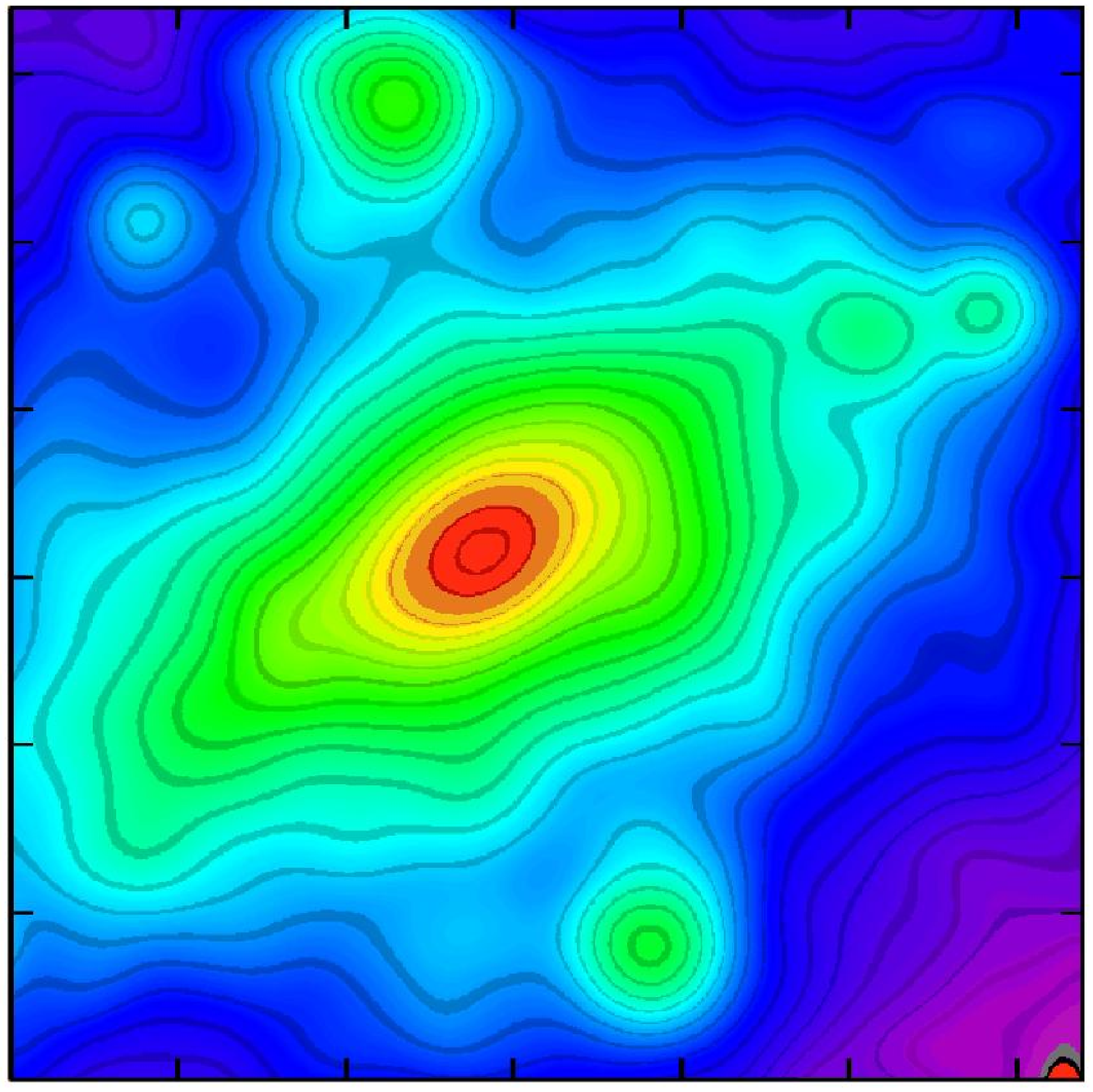}{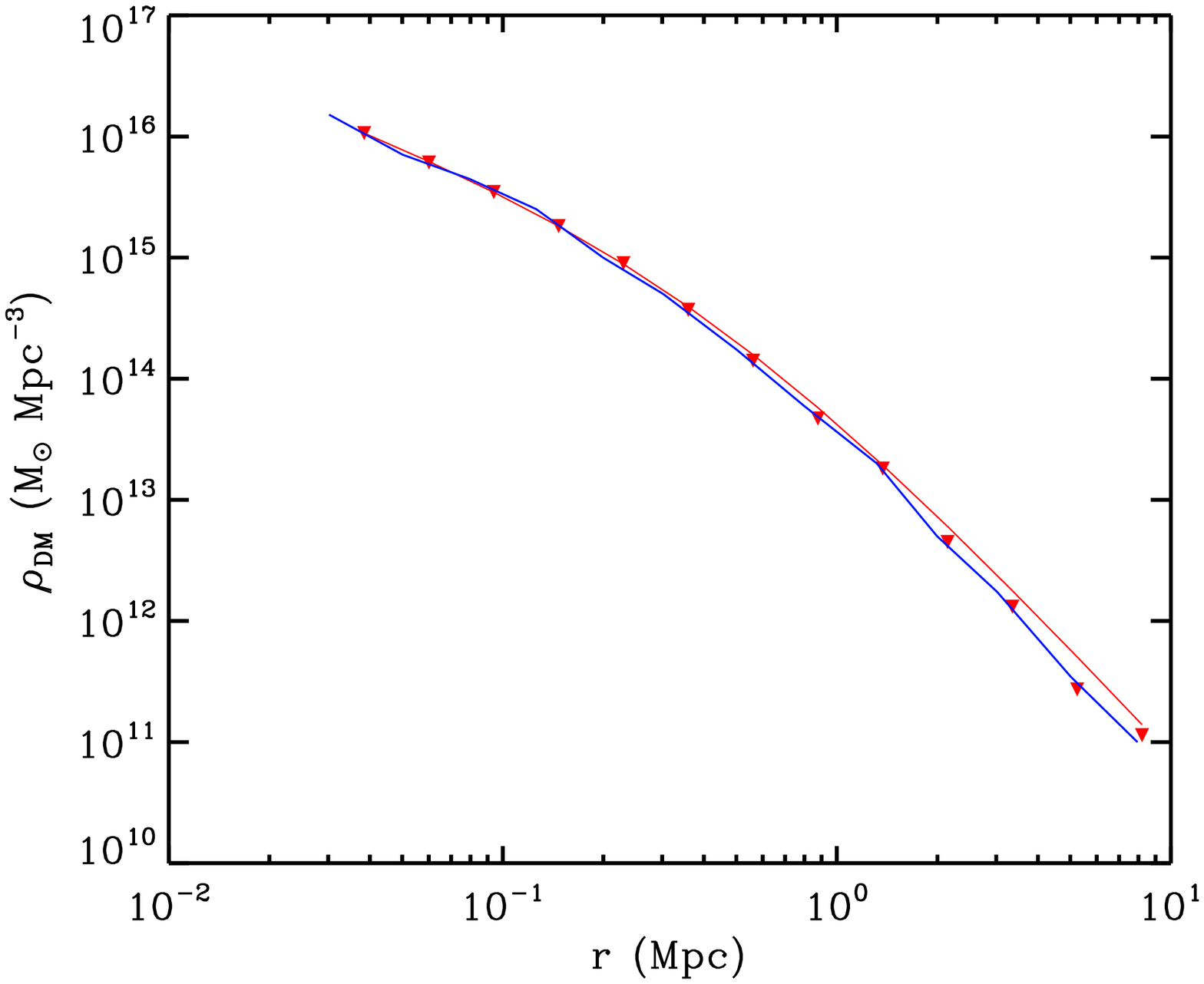}
\caption{Comparison with the Santa Barbara cluster
\citep{sta-barbara}.  The left panel shows the projected dark matter
density at $z=0$. The image, covering the inner 8~Mpc of the
simulation cube, has been smoothed with a Gaussian kernel of $R_{\rm
G} = 250$~kpc. The right panel shows the radial spherical density
distribution for the halo (red triangles), together with the mean
points from the Santa Barbara project (blue line). The red line
indicates the best NFW fit to the data points.}
\label{fig:sta-barbara}
\end{figure}


\section{Comparison of the FTM code with the Santa Barbara Cluster}

``The Santa Barbara cluster comparison project'' \citep{sta-barbara}
was envisioned as a reliable comparison test between different
cosmological numerical codes. It represents a hydrodynamical
simulation of the formation of a rich cluster of galaxies. This
exercise has become a standard test for new numerical codes in order
to check for broad consistencies with other codes. Here we limit
ourself to present only a comparison with respect to the DM properties
of such cluster. The FTM has been tested extensively with respect to
other dynamical codes.

We have simulated the Santa Barbara cluster with a numerical
resolution of $128^3$ particles with a homogeneous sampling from the
original $256^3$ realization. The left-panel of
Figure~\ref{fig:sta-barbara} shows a density cut of the final output
time from the simulation. This figure has been constructed in the same
way as those presented in \cite{sta-barbara}, \ie, the image covers
the inner 8~Mpc of the simulation cube, and it has been smoothed with
a Gaussian kernel of $R_{\rm G} = 250$~kpc. A visual comparison shows
that our simulation possesses the same characteristics as the Santa
Barbara one. This is further corroborated with the computed virial
quantities: $M_{\rm vir} = 1.06 \times 10^{15} ~\msun$, $R_{\rm vir} =
2.63$~Mpc, which are in good agreement with those reported in the
Santa Barbara project. In the right panel of
Fig.~\ref{fig:sta-barbara} the radial density distribution of our
cluster is presented. The red triangles represent our measurements and
the red line shows the corresponding NFW profile. This distribution is
in good agreement with respect to the averaged density profile
reported in the Santa Barbara project.

\bibliography{ms_2c}

\begin{thebibliography}{89}
\expandafter\ifx\csname natexlab\endcsname\relax\def\natexlab#1{#1}\fi

\bibitem[{{Allgood} {et~al.}(2006){Allgood}, {Flores}, {Primack}, {Kravtsov},
  {Wechsler}, {Faltenbacher}, \& {Bullock}}]{allgood06}
{Allgood}, B., {Flores}, R.~A., {Primack}, J.~R., {Kravtsov}, A.~V.,
  {Wechsler}, R.~H., {Faltenbacher}, A., \& {Bullock}, J.~S. 2006, \mnras, 367,
  1781

\bibitem[{{Ascasibar} {et~al.}(2004){Ascasibar}, {Yepes}, {Gottl{\"o}ber}, \&
  {M{\"u}ller}}]{ascasibar04}
{Ascasibar}, Y., {Yepes}, G., {Gottl{\"o}ber}, S., \& {M{\"u}ller}, V. 2004,
  \mnras, 352, 1109

\bibitem[{{Avila-Reese} {et~al.}(2005){Avila-Reese}, {Col{\'{\i}}n},
  {Gottl{\"o}ber}, {Firmani}, \& {Maulbetsch}}]{avila05}
{Avila-Reese}, V., {Col{\'{\i}}n}, P., {Gottl{\"o}ber}, S., {Firmani}, C., \&
  {Maulbetsch}, C. 2005, \apj, 634, 51

\bibitem[{{Barnes} \& {Efstathiou}(1987)}]{barnes87}
{Barnes}, J. \& {Efstathiou}, G. 1987, \apj, 319, 575

\bibitem[{{Barnes} \& {Hut}(1986)}]{Barnes:1986aa}
{Barnes}, J. \& {Hut}, P. 1986, \nat, 324, 446

\bibitem[{{Berentzen} \& {Shlosman}(2006)}]{ingo06}
{Berentzen}, I. \& {Shlosman}, I. 2006, \apj, 648, 807

\bibitem[{{Bertschinger}(1985)}]{bert85}
{Bertschinger}, E. 1985, \apjs, 58, 39

\bibitem[{{Bertschinger}(1987)}]{bert87}
---. 1987, \apjl, 323, L103

\bibitem[{{Binney}(2004)}]{Binney:2004aa}
{Binney}, J. 2004, \mnras, 350, 939

\bibitem[{{Bistolas} \& {Hoffman}(1998)}]{bh98}
{Bistolas}, V. \& {Hoffman}, Y. 1998, \apj, 492, 439

\bibitem[{{Blais-Ouellette} {et~al.}(2001){Blais-Ouellette}, {Amram}, \&
  {Carignan}}]{blais01}
{Blais-Ouellette}, S., {Amram}, P., \& {Carignan}, C. 2001, \aj, 121, 1952

\bibitem[{{Bolatto} {et~al.}(2002){Bolatto}, {Simon}, {Leroy}, \&
  {Blitz}}]{bolatto02}
{Bolatto}, A.~D., {Simon}, J.~D., {Leroy}, A., \& {Blitz}, L. 2002, \apj, 565,
  238

\bibitem[{{Broadhurst} {et~al.}(2005){Broadhurst}, {Takada}, {Umetsu}, {Kong},
  {Arimoto}, {Chiba}, \& {Futamase}}]{broadhurst05}
{Broadhurst}, T., {Takada}, M., {Umetsu}, K., {Kong}, X., {Arimoto}, N.,
  {Chiba}, M., \& {Futamase}, T. 2005, \apjl, 619, L143

\bibitem[{{Bullock} {et~al.}(2001{\natexlab{a}}){Bullock}, {Dekel}, {Kolatt},
  {Kravtsov}, {Klypin}, {Porciani}, \& {Primack}}]{bullock01ck}
{Bullock}, J.~S., {Dekel}, A., {Kolatt}, T.~S., {Kravtsov}, A.~V., {Klypin},
  A.~A., {Porciani}, C., \& {Primack}, J.~R. 2001{\natexlab{a}}, \apj, 555, 240

\bibitem[{{Bullock} {et~al.}(2001{\natexlab{b}}){Bullock}, {Kolatt}, {Sigad},
  {Somerville}, {Kravtsov}, {Klypin}, {Primack}, \& {Dekel}}]{bullock01sf}
{Bullock}, J.~S., {Kolatt}, T.~S., {Sigad}, Y., {Somerville}, R.~S.,
  {Kravtsov}, A.~V., {Klypin}, A.~A., {Primack}, J.~R., \& {Dekel}, A.
  2001{\natexlab{b}}, \mnras, 321, 559

\bibitem[{{Cole} \& {Lacey}(1996)}]{cl96}
{Cole}, S. \& {Lacey}, C. 1996, \mnras, 281, 716

\bibitem[{{Davis} {et~al.}(1985){Davis}, {Efstathiou}, {Frenk}, \&
  {White}}]{defw85}
{Davis}, M., {Efstathiou}, G., {Frenk}, C.~S., \& {White}, S.~D.~M. 1985, \apj,
  292, 371

\bibitem[{{de Blok} \& {Bosma}(2002)}]{de-Blok:2002aa}
{de Blok}, W.~J.~G. \& {Bosma}, A. 2002, \aap, 385, 816

\bibitem[{{Dehnen}(2002)}]{dehn02}
{Dehnen}, W. 2002, J. Comp. Phys., 27

\bibitem[{{Dehnen}(2005)}]{dehn05}
---. 2005, \mnras, 360, 892

\bibitem[{{Diemand} {et~al.}(2004){Diemand}, {Moore}, {Stadel}, \&
  {Kazantzidis}}]{diemand04}
{Diemand}, J., {Moore}, B., {Stadel}, J., \& {Kazantzidis}, S. 2004, \mnras,
  348, 977

\bibitem[{{Eisenstein} \& {Hut}(1998)}]{hop}
{Eisenstein}, D.~J. \& {Hut}, P. 1998, \apj, 498, 137

\bibitem[{{Eke} {et~al.}(2001){Eke}, {Navarro}, \& {Steinmetz}}]{ens01}
{Eke}, V.~R., {Navarro}, J.~F., \& {Steinmetz}, M. 2001, \apj, 554, 114

\bibitem[{{El-Zant}(2005)}]{ez05}
{El-Zant}, A. 2005, astro-ph/0502472

\bibitem[{{El-Zant} {et~al.}(2001){El-Zant}, {Shlosman}, \&
  {Hoffman}}]{El-Zant:2001aa}
{El-Zant}, A., {Shlosman}, I., \& {Hoffman}, Y. 2001, \apj, 560, 636

\bibitem[{{El-Zant} {et~al.}(2004){El-Zant}, {Hoffman}, {Primack}, {Combes}, \&
  {Shlosman}}]{El-Zant:2004aa}
{El-Zant}, A.~A., {Hoffman}, Y., {Primack}, J., {Combes}, F., \& {Shlosman}, I.
  2004, \apjl, 607, L75

\bibitem[{{Fillmore} \& {Goldreich}(1984)}]{fg84}
{Fillmore}, J.~A. \& {Goldreich}, P. 1984, \apj, 281, 1

\bibitem[{{Flores} \& {Primack}(1994)}]{fp94}
{Flores}, R.~A. \& {Primack}, J.~R. 1994, \apjl, 427, L1

\bibitem[{{Frenk} {et~al.}(1999){Frenk}, {White}, {Bode}, {Bond}, {Bryan},
  {Cen}, {Couchman}, {Evrard}, {Gnedin}, {Jenkins}, {Khokhlov}, {Klypin},
  {Navarro}, {Norman}, {Ostriker}, {Owen}, {Pearce}, {Pen}, {Steinmetz},
  {Thomas}, {Villumsen}, {Wadsley}, {Warren}, {Xu}, \& {Yepes}}]{sta-barbara}
{Frenk}, C.~S., {White}, S.~D.~M., {Bode}, P., {Bond}, J.~R., {Bryan}, G.~L.,
  {Cen}, R., {Couchman}, H.~M.~P., {Evrard}, A.~E., {Gnedin}, N., {Jenkins},
  A., {Khokhlov}, A.~M., {Klypin}, A., {Navarro}, J.~F., {Norman}, M.~L.,
  {Ostriker}, J.~P., {Owen}, J.~M., {Pearce}, F.~R., {Pen}, U.-L., {Steinmetz},
  M., {Thomas}, P.~A., {Villumsen}, J.~V., {Wadsley}, J.~W., {Warren}, M.~S.,
  {Xu}, G., \& {Yepes}, G. 1999, \apj, 525, 554

\bibitem[{{Fukushige} \& {Makino}(1997)}]{fm97}
{Fukushige}, T. \& {Makino}, J. 1997, \apjl, 477, L9+

\bibitem[{{Fukushige} \& {Makino}(2003)}]{fm03}
---. 2003, \apj, 588, 674

\bibitem[{{Gardner}(2001)}]{gardner01}
{Gardner}, J.~P. 2001, \apj, 557, 616

\bibitem[{{Gelb} \& {Bertschinger}(1994)}]{denmax}
{Gelb}, J.~M. \& {Bertschinger}, E. 1994, \apj, 436, 467

\bibitem[{{Gentile} {et~al.}(2004){Gentile}, {Salucci}, {Klein}, {Vergani}, \&
  {Kalberla}}]{gen04}
{Gentile}, G., {Salucci}, P., {Klein}, U., {Vergani}, D., \& {Kalberla}, P.
  2004, \mnras, 351, 903

\bibitem[{{Ghigna} {et~al.}(2000){Ghigna}, {Moore}, {Governato}, {Lake},
  {Quinn}, \& {Stadel}}]{ghigna00}
{Ghigna}, S., {Moore}, B., {Governato}, F., {Lake}, G., {Quinn}, T., \&
  {Stadel}, J. 2000, \apj, 544, 616

\bibitem[{{Gunn}(1977)}]{gunn77}
{Gunn}, J.~E. 1977, \apj, 218, 592

\bibitem[{{Gunn} \& {Gott}(1972)}]{gg72}
{Gunn}, J.~E. \& {Gott}, J.~R.~I. 1972, \apj, 176, 1

\bibitem[{{Heller}(1995)}]{hel95}
{Heller}, C.~H. 1995, \apj, 455, 252

\bibitem[{{Heller} \& {Shlosman}(1994)}]{hel94}
{Heller}, C.~H. \& {Shlosman}, I. 1994, \apj, 424, 84

\bibitem[{{Hernquist}(1990)}]{h90}
{Hernquist}, L. 1990, \apj, 356, 359

\bibitem[{{Hiotelis}(2002)}]{Hiotelis:2002aa}
{Hiotelis}, N. 2002, \aap, 382, 84

\bibitem[{{Hoekstra} {et~al.}(2004){Hoekstra}, {Yee}, \&
  {Gladders}}]{Hoekstra:2004aa}
{Hoekstra}, H., {Yee}, H.~K.~C., \& {Gladders}, M.~D. 2004, \apj, 606, 67

\bibitem[{{Hoffman} \& {Ribak}(1991)}]{hr91}
{Hoffman}, Y. \& {Ribak}, E. 1991, \apjl, 380, L5

\bibitem[{{Hoffman} \& {Shaham}(1985)}]{hs85}
{Hoffman}, Y. \& {Shaham}, J. 1985, \apj, 297, 16

\bibitem[{{Jing} \& {Suto}(2000)}]{js00}
{Jing}, Y.~P. \& {Suto}, Y. 2000, \apjl, 529, L69

\bibitem[{{Jing} \& {Suto}(2002)}]{js02}
---. 2002, \apj, 574, 538

\bibitem[{{Klypin} {et~al.}(2001){Klypin}, {Kravtsov}, {Bullock}, \&
  {Primack}}]{kkbp01}
{Klypin}, A., {Kravtsov}, A.~V., {Bullock}, J.~S., \& {Primack}, J.~R. 2001,
  \apj, 554, 903

\bibitem[{{Kravtsov} {et~al.}(2002){Kravtsov}, {Klypin}, \& {Hoffman}}]{kkh02}
{Kravtsov}, A.~V., {Klypin}, A., \& {Hoffman}, Y. 2002, \apj, 571, 563

\bibitem[{{Lemson} \& {Kauffmann}(1999)}]{lk99}
{Lemson}, G. \& {Kauffmann}, G. 1999, \mnras, 302, 111

\bibitem[{{{\L}okas} \& {Hoffman}(2000)}]{lh00}
{{\L}okas}, E.~L. \& {Hoffman}, Y. 2000, \apjl, 542, L139

\bibitem[{{Lu} {et~al.}(2006){Lu}, {Mo}, {Katz}, \& {Weinberg}}]{Lu:2006aa}
{Lu}, Y., {Mo}, H.~J., {Katz}, N., \& {Weinberg}, M.~D. 2006, \mnras, 368, 1931

\bibitem[{{Macci{\`o}} {et~al.}(2003){Macci{\`o}}, {Murante}, \&
  {Bonometto}}]{mmb03}
{Macci{\`o}}, A.~V., {Murante}, G., \& {Bonometto}, S.~P. 2003, \apj, 588, 35

\bibitem[{{Moore} {et~al.}(1998){Moore}, {Governato}, {Quinn}, {Stadel}, \&
  {Lake}}]{moore98}
{Moore}, B., {Governato}, F., {Quinn}, T., {Stadel}, J., \& {Lake}, G. 1998,
  \apjl, 499, L5+

\bibitem[{{Moore} {et~al.}(1999){Moore}, {Quinn}, {Governato}, {Stadel}, \&
  {Lake}}]{moore99}
{Moore}, B., {Quinn}, T., {Governato}, F., {Stadel}, J., \& {Lake}, G. 1999,
  \mnras, 310, 1147

\bibitem[{{Navarro} {et~al.}(1997){Navarro}, {Frenk}, \& {White}}]{nfw97}
{Navarro}, J.~F., {Frenk}, C.~S., \& {White}, S.~D.~M. 1997, \apj, 490, 493

\bibitem[{{Nusser}(2001)}]{nusser01}
{Nusser}, A. 2001, \mnras, 325, 1397

\bibitem[{{Nusser} \& {Sheth}(1999)}]{ns99}
{Nusser}, A. \& {Sheth}, R.~K. 1999, \mnras, 303, 685

\bibitem[{{Padmanabhan}(1993)}]{p93}
{Padmanabhan}, T. 1993, {Structure Formation in the Universe} (Structure
  Formation in the Universe, by T.~Padmanabhan, pp.~499.~ISBN
  0521424860.~Cambridge, UK: Cambridge University Press, June 1993.)

\bibitem[{{Peebles}(1969)}]{peebles69}
{Peebles}, P.~J.~E. 1969, \apj, 155, 393

\bibitem[{{Peirani} {et~al.}(2004){Peirani}, {Mohayaee}, \& {de Freitas
  Pacheco}}]{peirani04}
{Peirani}, S., {Mohayaee}, R., \& {de Freitas Pacheco}, J.~A. 2004, \mnras,
  348, 921

\bibitem[{{Reed} {et~al.}(2005{\natexlab{a}}){Reed}, {Governato}, {Quinn},
  {Gardner}, {Stadel}, \& {Lake}}]{Reed:2005iv}
{Reed}, D., {Governato}, F., {Quinn}, T., {Gardner}, J., {Stadel}, J., \&
  {Lake}, G. 2005{\natexlab{a}}, \mnras, 359, 1537

\bibitem[{{Reed} {et~al.}(2005{\natexlab{b}}){Reed}, {Governato}, {Verde},
  {Gardner}, {Quinn}, {Stadel}, {Merritt}, \& {Lake}}]{Reed:2005aa}
{Reed}, D., {Governato}, F., {Verde}, L., {Gardner}, J., {Quinn}, T., {Stadel},
  J., {Merritt}, D., \& {Lake}, G. 2005{\natexlab{b}}, \mnras, 357, 82

\bibitem[{{Ricotti}(2003)}]{ricotti03}
{Ricotti}, M. 2003, \mnras, 344, 1237

\bibitem[{{Romano-D{\'{\i}}az}(2004)}]{erd04}
{Romano-D{\'{\i}}az}, E. 2004, PhD thesis, University of Groningen, Kapteyn
  Astronomical Institute

\bibitem[{{Romano-D{\'{\i}}az} {et~al.}(2006){Romano-D{\'{\i}}az},
  {Faltenbacher}, {Jones}, {Heller}, {Hoffman}, \& {Shlosman}}]{erd06}
{Romano-D{\'{\i}}az}, E., {Faltenbacher}, A., {Jones}, D., {Heller}, C.,
  {Hoffman}, Y., \& {Shlosman}, I. 2006, \apjl, 637, L93

\bibitem[{{Ryden}(1988)}]{ryden88}
{Ryden}, B.~S. 1988, \apj, 329, 589

\bibitem[{{Ryden} \& {Gunn}(1987)}]{rg87}
{Ryden}, B.~S. \& {Gunn}, J.~E. 1987, \apj, 318, 15

\bibitem[{{Salucci} \& {Burkert}(2000)}]{sb00}
{Salucci}, P. \& {Burkert}, A. 2000, \apjl, 537, L9

\bibitem[{{Shaw} {et~al.}(2005){Shaw}, {Weller}, {Ostriker}, \&
  {Bode}}]{Shaw:2005ua}
{Shaw}, L., {Weller}, J., {Ostriker}, J.~P., \& {Bode}, P. 2005,
  astro-ph/0509856

\bibitem[{{Simon} {et~al.}(2003){Simon}, {Bolatto}, {Leroy}, \&
  {Blitz}}]{simon03}
{Simon}, J.~D., {Bolatto}, A.~D., {Leroy}, A., \& {Blitz}, L. 2003, \apj, 596,
  957

\bibitem[{{Simon} {et~al.}(2005){Simon}, {Bolatto}, {Leroy}, {Blitz}, \&
  {Gates}}]{simon05}
{Simon}, J.~D., {Bolatto}, A.~D., {Leroy}, A., {Blitz}, L., \& {Gates}, E.~L.
  2005, \apj, 621, 757

\bibitem[{{Springel} {et~al.}(2001){Springel}, {White}, {Tormen}, \&
  {Kauffmann}}]{subfind}
{Springel}, V., {White}, S.~D.~M., {Tormen}, G., \& {Kauffmann}, G. 2001,
  \mnras, 328, 726

\bibitem[{{Steinmetz} \& {Bartelmann}(1995)}]{sb95}
{Steinmetz}, M. \& {Bartelmann}, M. 1995, \mnras, 272, 570

\bibitem[{{Subramanian} {et~al.}(2000){Subramanian}, {Cen}, \&
  {Ostriker}}]{sco00}
{Subramanian}, K., {Cen}, R., \& {Ostriker}, J.~P. 2000, \apj, 538, 528

\bibitem[{{Syer} \& {White}(1998)}]{sw98}
{Syer}, D. \& {White}, S.~D.~M. 1998, \mnras, 293, 337

\bibitem[{{Tasitsiomi} {et~al.}(2004){Tasitsiomi}, {Kravtsov}, {Gottl{\"o}ber},
  \& {Klypin}}]{t04}
{Tasitsiomi}, A., {Kravtsov}, A.~V., {Gottl{\"o}ber}, S., \& {Klypin}, A.~A.
  2004, \apj, 607, 125

\bibitem[{{Taylor} \& {Navarro}(2001)}]{tn01}
{Taylor}, J.~E. \& {Navarro}, J.~F. 2001, \apj, 563, 483

\bibitem[{{van de Weygaert} \& {Bertschinger}(1996)}]{vdw96}
{van de Weygaert}, R. \& {Bertschinger}, E. 1996, \mnras, 281, 84

\bibitem[{{van den Bosch}(2002)}]{vdb02}
{van den Bosch}, F.~C. 2002, \mnras, 331, 98

\bibitem[{{Vitvitska} {et~al.}(2002){Vitvitska}, {Klypin}, {Kravtsov},
  {Wechsler}, {Primack}, \& {Bullock}}]{v02}
{Vitvitska}, M., {Klypin}, A.~A., {Kravtsov}, A.~V., {Wechsler}, R.~H.,
  {Primack}, J.~R., \& {Bullock}, J.~S. 2002, \apj, 581, 799

\bibitem[{{Warren} {et~al.}(1992){Warren}, {Quinn}, {Salmon}, \& {Zurek}}]{w92}
{Warren}, M.~S., {Quinn}, P.~J., {Salmon}, J.~K., \& {Zurek}, W.~H. 1992, \apj,
  399, 405

\bibitem[{{Wechsler} {et~al.}(2002){Wechsler}, {Bullock}, {Primack},
  {Kravtsov}, \& {Dekel}}]{w02}
{Wechsler}, R.~H., {Bullock}, J.~S., {Primack}, J.~R., {Kravtsov}, A.~V., \&
  {Dekel}, A. 2002, \apj, 568, 52

\bibitem[{{Wechsler} {et~al.}(2005){Wechsler}, {Zentner}, {Bullock},
  {Kravtsov}, \& {Allgood}}]{w05}
{Wechsler}, R.~H., {Zentner}, A.~R., {Bullock}, J.~S., {Kravtsov}, A.~V., \&
  {Allgood}, B. 2005, astro-ph/0512416

\bibitem[{{Weinberg} \& {Katz}(2002)}]{Weinberg:2002aa}
{Weinberg}, M.~D. \& {Katz}, N. 2002, \apj, 580, 627

\bibitem[{{Weldrake} {et~al.}(2003){Weldrake}, {de Blok}, \&
  {Walter}}]{weldrake03}
{Weldrake}, D.~T.~F., {de Blok}, W.~J.~G., \& {Walter}, F. 2003, \mnras, 340,
  12

\bibitem[{{Zaroubi} \& {Hoffman}(1993)}]{zh93}
{Zaroubi}, S. \& {Hoffman}, Y. 1993, \apj, 416, 410

\bibitem[{{Zel'dovich}(1970)}]{zeldovich}
{Zel'dovich}, Y.~B. 1970, \aap, 5, 84

\bibitem[{{Zhao} {et~al.}(2003){Zhao}, {Mo}, {Jing}, \& {B{\"o}rner}}]{z03b}
{Zhao}, D.~H., {Mo}, H.~J., {Jing}, Y.~P., \& {B{\"o}rner}, G. 2003, \mnras,
  339, 12

\bibitem[{{Zhao}(1996)}]{zhao96}
{Zhao}, H. 1996, \mnras, 278, 488

\end{thebibliography}


\end{document}